\documentclass[twocolumn]{aastex7}
\usepackage{epsfig}
\usepackage{graphicx}
\usepackage{float}
\usepackage{amsmath}
\usepackage{color} 
\usepackage{amssymb}
\usepackage{amsfonts}
\usepackage{units}
\usepackage{bm}
\usepackage{longtable} 
\usepackage{booktabs}
\usepackage{multirow}
\usepackage[mathscr]{euscript}
\usepackage{soul}

\newcommand\be{\begin{eqnarray}}
\newcommand\ee{\end{eqnarray}}

\newcommand{\fraction}[3]{\left(\frac{#1}{#2}\right)^{#3}}
\newcommand{\fractionz}[2]{\left(\frac{#1}{#2}\right)} 

\begin{document}

\title{Magnetic White Dwarf -- M Dwarf Binaries in Pre-mCV Phase as Special Population of Long-Period Radio Transients}

\correspondingauthor{Yuan-Pei Yang (ypyang@ynu.edu.cn)}

\author[0000-0001-6374-8313]{Yuan-Pei Yang}
\affiliation{South-Western Institute for Astronomy Research, Yunnan Key Laboratory of Survey Science, Yunnan University, Kunming, Yunnan 650504, People's Republic of China}
\email{ypyang@ynu.edu.cn}  

\begin{abstract}

Long-period radio transients (LPTs) are a new class of coherent radio sources with periods ranging from minutes to hours. Recently, two LPT sources, ILT J1101+5521 and GLEAM-X J0704-37, with periods of 2-3 hours has been confirmed to originate from white dwarf (WD) -- M dwarf (MD) binaries. In this work, we propose that at least some LPTs originate from the magnetic WD -- MD binaries in the pre-magnetic cataclysmic variables (pre-mCV) phase. The asynchronism between the WD's rotation and the binary's orbital motion allows for the unipolar-inductor mechanism or magnetosphere interaction to operate and accelerate radiating particles, with the dominant process depending on the magnetic moment ratio of the two stars. Under asynchronism condition, both the peak flux and the polarization of radio pulses will be modulated by the beat period. The pre-mCV phase characterized by an extremely low accretion rate provides the relatively clean magnetospheric environment necessary for a loss-cone-driven maser (LCDM) mechanism to operate, producing the LPT emission. The observed pulse duty cycle of $10^{-3}-10^{-1}$ is attributed to a beaming effect modulated by the binary's magnetic geometry. Furthermore, the magnetized environment of a WD--MD binary is conducive to Faraday conversion with weak coupling, which implies that the polarization state of LPTs should vary significantly at different periods. Finally, we predict that LPTs from WD--MD binaries should exhibit a period distribution following $f_P(P)dP \propto P^{(1.67-2.33)}dP$ and a luminosity function described by $f_L(L)dL \propto L^{-(1.80-2.67)}dL$, which can be tested by the future large sample. 

\end{abstract}


\section{Introduction}  

Long-period radio transients (LPTs) are an emerging class of coherent radio sources characterized by periods ranging from minutes to several hours, placing them beyond the ``death line'' for radio pulsars. The observed activity timescales of these sources are diverse; some are transient, remaining active for only a few months, whereas others appear to be long-lived, with emission persisting over several years of observation. The emission from LPTs typically consists of intense, coherent radio pulses with duration of tens to hundreds of seconds, which exhibit significant brightness variability. Furthermore, the individual pulses often display complex morphologies, including features such as drifting substructures and quasi-periodic oscillations \citep{Hurley-Walker23,Men25}. 
A key diagnostic feature of LPTs is the high degree of polarization in their radio emission. The polarization properties are varied: some sources, such as GPM J1839-10 and ASKAP J1839-0756, exhibit a large fraction of linear polarization (from 10\% to 100\%) \citep{Hurley-Walker23,Lee25}, while others, like CHIME J1634+44, have been observed to emit nearly fully circularly polarized ($>90\%$) bursts \citep{Dong25}. The polarization state can also vary dramatically within a single source. A notable example is ASKAP J1935+2148, which, with a period of 53.8 min, displays three distinct emission states: a bright-pulse state with highly linear polarization, a weak-pulse state with highly circular polarization, and a quiescent state with no detectable pulses \citep{Caleb24}. In addition, sources like GPM J1839-10 and ASKAP J1839-0756 exhibit complex phenomena such as polarization angle swings, orthogonal polarization mode jumps, and Faraday conversion \citep{Men25,Hurley-Walker23,Lee25}. These intricate behaviors underscore the enigmatic nature of LPTs and establish them as a compelling subject for contemporary astrophysical research.

The physical nature of LPTs remains a subject of active debate. Proposed theoretical models can be broadly divided into two main classes based on the mechanism driving the periodicity: (i) the rotation of a compact object, such as a neutron star or a white dwarf (WD) \citep{Katz22,Beniamini,Rea24,Cary25}; and (ii) the orbital motion of a WD--M dwarf (MD) binary system \citep{deRuiter25,Hurley-Walker24,Rodriguez25,Qu25,Horvath25}. Both scenarios are supported by compelling, albeit distinct, lines of observational evidence.
Support for the neutron star rotation model comes from several key observations. For instance, the LPT known as ASKAP/DART-J1832 has been found to be spatially coincident with the supernova remnant G22.7-0.2 \citep{Li24}, and pulsed X-ray emission was detected from its position \citep{Wang25}. Furthermore, the magnetar candidate IE 161348-5055, located in the supernova remnant RCW 103, exhibits a long period of 6.7 hours, although radio emission has not yet been detected \citep{DeLuca06}. Collectively, these findings suggest that slowly rotating neutron stars are a plausible origin for a subset of the LPT population.
Conversely, strong evidence for the binary model has recently emerged from observations of two specific LPTs. ILT J1101+5521 (with a period of 125.5 min) and GLEAM-X J0704-37 (with a period of 174 min) have been definitively linked to WD--MD binary systems \citep{deRuiter25,Hurley-Walker24,Rodriguez25}. This association was established through optical spectroscopy of the MD companions, which revealed orbital periods that well match the radio periodicities of the LPTs. These direct associations provide unambiguous proof that at least some LPTs originate from WD--MD binaries. Therefore, the emerging consensus is that the LPT population likely has multiple origins. 

\begin{figure}
    \centering
    \includegraphics[width = 1.0\linewidth, trim = 0 0 0 0, clip]{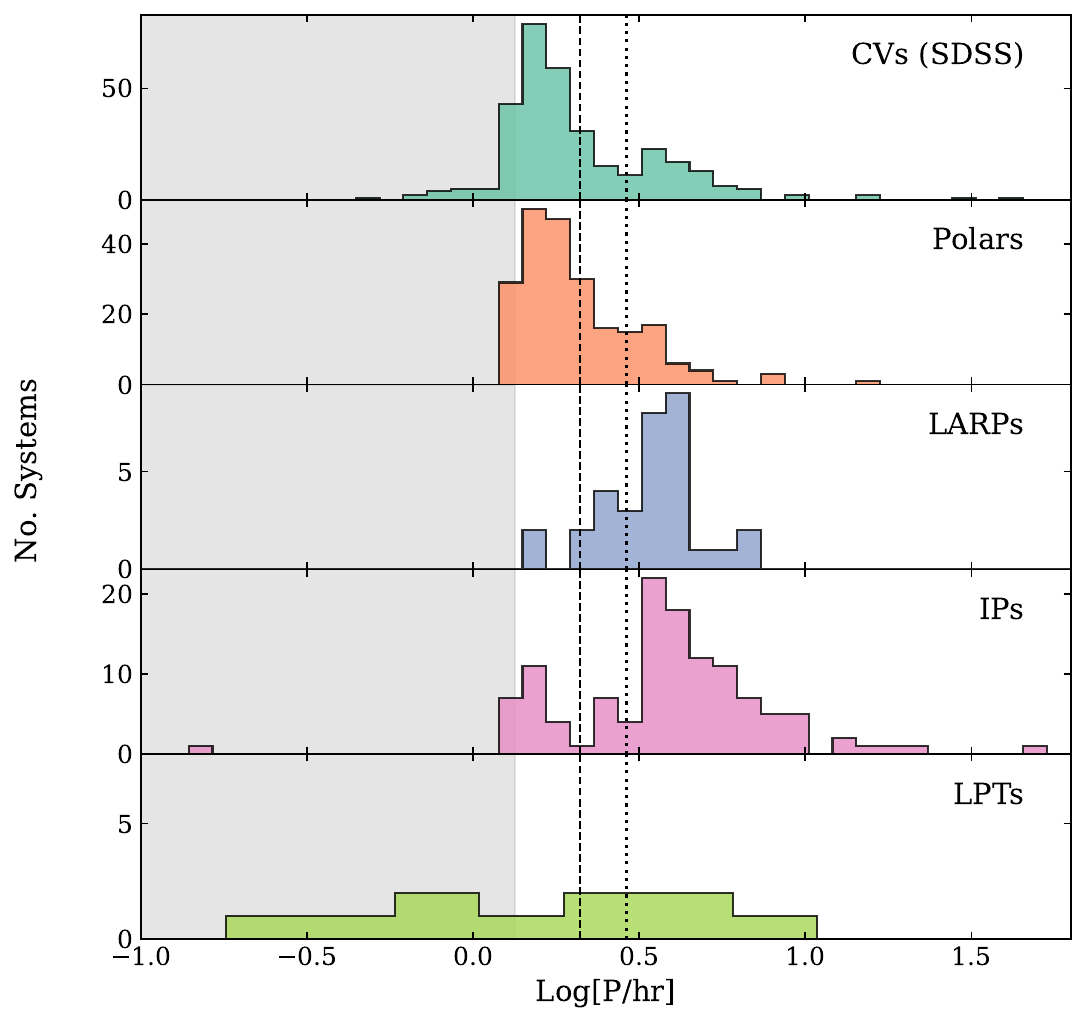} 
    \caption{Period distributions of LPTs shown alongside populations of CVs and mCVs (including Polars, LARPs, and IPs). The population data are compiled as follows: CVs from the SDSS I--IV archives \citep{Inight23}; Polars and LARPs from the PolarCat catalog \citep{Schwope25}; IPs from the online catalog maintained by Koji Mukai's ``The Intermediate Polars'' homepage (\url{https://asd.gsfc.nasa.gov/Koji.Mukai/iphome/iphome.html}, also see \citet{Mukai17}). The data of LPTs is from Table 1 of \citet{Qu25}. The vertical dashed and dotted lines mark the periods of the confirmed WD--MD binary LPTs, ILT J1101+5521 (125.5 min) and GLEAM-X J0704-37 (174 min), respectively. The grey shaded region indicates periods below the canonical $\sim$80-minute orbital minimum for CVs/mCVs.
    }\label{CVperiod}
\end{figure}

WD--MD binary systems are crucial astrophysical laboratories, responsible for a diverse array of energetic phenomena. The specific manifestation depends on key physical parameters, such as the binary separation and the strength of the magnetic interaction. These systems give rise to various phenomena, including classical cataclysmic variables (CVs), which exhibit novae and dwarf novae outbursts, and magnetic cataclysmic variables (mCVs), a category that includes Polars, low-accretion rate polars (LARPs; also known as ``pre-polars''), and intermediate polars (IPs). 
Crucially, the orbital periods of these close binary systems overlap with the range observed for LPTs, providing a natural basis for the binary origin. 
This overlap is illustrated in Figure \ref{CVperiod}, which displays the period distributions for these populations alongside those of the known LPTs. Notably, the confirmed WD--MD binary LPTs, ILT J1101+5521 and GLEAM-X J0704-37, fall squarely within this period range, near the classical ``period gap'' observed in CV populations. 
However, the binary origin can not account for the entire LPT population. Approximately half of the known LPTs exhibit periods significantly shorter than the canonical $\sim$80-minute orbital period minimum for CVs and mCVs (indicated by the grey region in Figure \ref{CVperiod}). This discrepancy strongly suggests an alternative origin for these shorter-period systems and reinforces the hypothesis that they may result from the rotation of neutron stars or WDs. 

In this work, we propose a model wherein a significant subset of LPTs originates from magnetic WD--MD binaries during their pre-mCV evolutionary phase. The pre-mCV phase here is generally defined as the evolutionary stage preceding the formation of mCV (observed Polars and IPs). Since the origin of magnetic fields in magnetic WDs remains a subject of debate \citep{Braithwaite04,Tout08,Schreiber21}, different hypotheses imply distinct evolutionary contexts for the pre-mCV stage. In scenarios invoking the fossil field at birth \citep{Braithwaite04} or the dynamo action during common-envelope evolution \citep{Tout08}, the WD magnetic field is generated relatively early (i.e., at the WD's formation or during common-envelope evolution), so the pre-mCV phase corresponds specifically to the pre-accretion stage of a magnetic WD--MD binary.  
Conversely, under the crystallization- and rotation-driven dynamo hypothesis \citep{Schreiber21}, magnetic field generation occurs later when the core of a rapidly rotating WD begins to crystallize in a WD--MD binary system. In this case, the pre-mCV phase represents a critical transitional stage in the evolution from CVs to mCVs\footnote{In this work, the term ``pre-mCV'' under the crystallization- and rotation-driven dynamo hypothesis \citep{Schreiber21} refers broadly to the evolutionary stage between the non-magnetic CV and fully-established mCV accretion states. This encompasses the initial magnetic field generation, a subsequent AR Sco-like phase, and the LARP phase.}: the newly formed magnetosphere disrupts ongoing mass transfer, effectively halting Roche-lobe overflow and producing a detached binary configuration. 
This stage thus marks the period after magnetic field emergence but before orbital shrinkage re-establishes significant accretion. 
Therefore, regardless of the magnetic field generation mechanisms, the accretion rate during the pre-mCV phase should remain low; with the MD companion underfilling its Roche lobe, mass transfer occurs only via a weak stellar wind, corresponding to a mass-loss rate of $\lesssim 10^{-14}~M_\odot~\mathrm{yr^{-1}}$ \citep{Wood05}.

Within this framework, these LPTs serve not only as novel phenomena but also as crucial probes of the astrophysical processes during this stage. We develop a comprehensive physical model to account for the key observed properties of these LPTs, including their energy dissipation mechanisms, radiation processes, propagation effects, and the statistical distribution of the population.
The paper is organized as follows. In Section~\ref{stage}, we first evaluate potential LPT generation scenarios throughout the evolution of a magnetic WD--MD binary, examining the accretion phase (Section~\ref{stage1}), the unipolar-inductor phase (Section~\ref{stage2}), and the magnetosphere-interaction phase (Section~\ref{stage3}). We then delve into the physics of the radiation process in Section~\ref{radiation}, where we discuss the geometry of the emission region (Section~\ref{beaming}), and analyze the loss-cone-driven maser (LCDM) mechanism as the most plausible emission process (Section~\ref{mechanism}). In Section~\ref{conversion}, we propose a model for Faraday conversion to explain the complex polarization evolution observed in LPTs. Subsequently, we discuss the statistical properties of the LPT population arising from WD--MD binaries and the cooling issue of the magnetic WDs in Section~\ref{population}. Finally, the results are discussed and summarized in Section \ref{discussion}.  

\section{Evaluating LPT Production Scenarios in WD--MD Binaries}\label{stage}  

While radial velocity measurements have confirmed that ILT J1101+5521 and GLEAM-X J0704-37 are associated with WD--MD binary systems \citep{deRuiter25,Hurley-Walker24,Rodriguez25}, their precise evolutionary stage remains uncertain. The majority of such binaries, notably CVs and mCVs, are observed in accretion-dominated phases characterized by outbursts and prominent X-ray/optical emission. However, a subset without significant accretion, e.g., pre-mCVs (including LARPs and AR Sco-like systems), has also been identified. 
This distinction is critical, as studies of analogous systems like millisecond pulsars in low-mass X-ray binaries (LMXBs) suggest that intense accretion generally suppresses the coherent radio emission required to produce LPTs (see Section~\ref{stage1} for further discussion).
To determine the most plausible origin for LPTs within this binary context, we will discuss three distinct evolutionary phases: the accretion phase, the unipolar-inductor phase, and the magnetosphere-interaction phase. Each stage is characterized by a different dominant energy dissipation mechanism and a distinct magnetospheric plasma environment. 
We define our system as a magnetic WD--MD binary with an orbital semi-major axis $a$ and angular velocity $\Omega$ (corresponding to an orbital period $P$). Given the short periods of interest, we assume the orbit is circular ($e \rightarrow 0$) due to long-term evolution driven by tidal forces, magnetic braking, mass transfer, etc. The system comprises a magnetic WD with mass $M_s$, radius $R_s$, surface magnetic field $B_s$, and spin angular velocity $\Omega_s$, alongside an MD companion with corresponding parameters $M_c$, $R_c$, and $B_c$. Their respective magnetic dipole moments are defined as $\mu_s = B_s R_s^3$ and $\mu_c = B_c R_c^3$.

\begin{figure}
    \centering
    \includegraphics[width = 1.0\linewidth, trim = 50 50 50 50, clip]{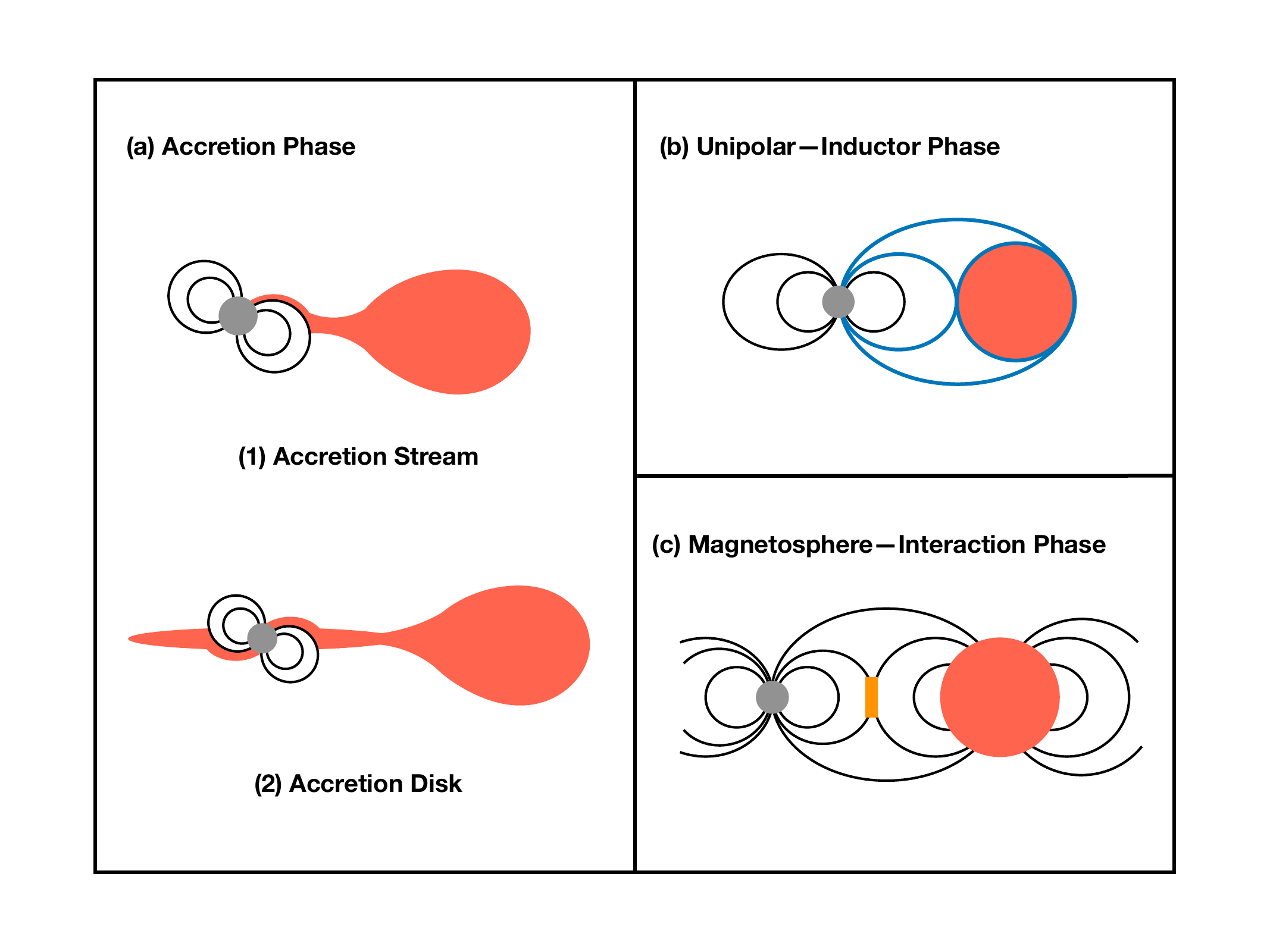} 
    \caption{Schematic illustration of the three primary evolutionary stages in a magnetic white dwarf (WD)--M dwarf (MD) binary. The panels depict: (a) the accretion phase, characterized by an accretion stream or an accretion disk around the WD; (b) the unipolar-inductor phase; and (c) the magnetosphere-interaction phase, where direct interaction between the two magnetospheres leads to magnetic reconnection (highlighted in orange).}\label{threemodel} 
\end{figure}

For ILT J1101+5521 and GLEAM-X J0704-37, their observed pulse periods of 125.5 min and 174 min have been established as the orbital period \citep{deRuiter25,Rodriguez25,Hurley-Walker24}. Thus, the semimajor axis can be estimated as
\begin{align}
a&=\left[\frac{G(M_s+M_c)P^2}{4\pi^2}\right]^{1/3}\simeq4.9\times10^{10}~{\rm cm}
\nonumber\\ 
&\times\fraction{P}{100~{\rm min}}{2/3}\fraction{M_s+M_c}{M_\odot}{1/3}.\label{Kepler3}
\end{align}
Typically, WDs span a mass range of $M_s\sim(0.2-1.4)M_\odot$, with the majority concentrated between $M_s\sim(0.5-0.7)M_\odot$ \citep[e.g.,][]{Gianninas10,Tremblay11}. Their MD companions have masses $M_c$ in the range of $M_c\sim (0.1-0.6)M_\odot$, with a distribution that peaks at $M_c\sim (0.1-0.3)M_\odot$ \citep[e.g.,][]{Chabrier03,Bochanski10}. The radius of the MD, $R_c$, is determined by its mass through the approximate relation \citep{Neece84,Caillault90}:
\be 
R_c\simeq 0.9 R_\odot \fraction{M_c}{M_\odot}{0.8}.\label{RM_relation}
\ee 
For $M_c\sim(0.1-0.6)M_\odot$, the radius range of MDs is $R_c\sim(1.0-4.2)\times10^{10}~{\rm cm}$. 
The estimated orbital semi-major axis of Eq.(\ref{Kepler3}) is comparable in magnitude to the intrinsic radius of the MD before accretion. This proximity allows the binary system to occupy several distinct evolutionary states, contingent upon specific orbital and stellar parameters. 
Both accretion-dominated phases (Roche-lobe overflow) and non-accreting phases (without Roche-lobe overflow) are therefore plausible, see Figure \ref{threemodel}. The latter encompasses the unipolar-inductor and magnetosphere-interaction scenarios. Next, we will analyze these three primary evolutionary stages in detail.

1) Accretion phase: 
Significant accretion in a WD--MD binary is driven by Roche-lobe overflow. This process is initiated when the MD expands to overfill its Roche lobe, causing matter to spill towards the WD,
\begin{align}
R_{\rm L1}=\eta a\lesssim R_c~{\rm with}~\eta 
\simeq 0.462\fraction{q}{1+q}{1/3},\label{roche} 
\end{align}
where $q=M_c/M_s$ is the mass ratio of the MD to the WD. 
The accuracy of this approximation is better than $4$\% for $0.1<q<1$  \citep{Paczynski71}. 
For WD -- MD binaries, the mass ratio is typically $q\sim(0.1-1)$, yielding a parameter range of $\eta\sim(0.2-0.4)$. 
If the binary system is in an accretion phase ($R_c \gtrsim R_{\rm L1}$), its orbital period must be
\begin{align} 
P&\lesssim P_{\rm acc}\equiv\left[\frac{4\pi^2 R_c^3}{G(M_s+M_c)\eta^3}\right]^{1/2}\simeq55~{\rm min}~\nonumber\\ 
&\times\fraction{\eta}{0.3}{-3/2}\fraction{M_s+M_c}{M_\odot}{-1/2}\fraction{R_c}{10^{10}~{\rm cm}}{3/2}.\label{Pacc} 
\end{align} 
Using the mass-radius relation of Eq.(\ref{RM_relation}) for MDs, the above equation can be rewritten as
\be 
P\lesssim P_{\rm acc}\simeq147~{\rm min}~\fraction{M_c}{0.2M_\odot}{0.7}\label{Pacc1}.
\ee
Notably, within Paczynski's approximation, this critical period is independent of the mass ratio $q$. It therefore represents an upper limit on the orbital period for any system in an accretion phase, a limit determined solely by the mass of the MD for $0.1\lesssim q \lesssim 1$.

The evolution of WD--MD binaries is also constrained by a minimum orbital period, $P_{\min} \sim 80$ min (indicated by the grey region in Figure~\ref{CVperiod}). This period minimum arises when continuous mass loss drives the MD to become less massive, leading to degeneration. At this point, its mass-radius relationship inverts to $R_c \propto M_c^{-1/3}$. Consequently, any subsequent mass loss causes the star to expand rather than contract. This stellar expansion, in turn, drives the binary into a wider orbit, a phenomenon known as the ``period bounce'' \citep{Paczynski81,Rappaport83}. Therefore, for a system to sustain accretion, its orbital period must lie within the bounds defined by this minimum and the previously derived accretion limit: $P_{\min} \lesssim P \lesssim P_{\rm acc}$.

2) Unipolar-inductor phase: 
When the orbital separation is sufficiently large such that the MD underfills its Roche lobe ($a > R_c/\eta$), accretion via Roche-lobe overflow ceases. The system enters a detached state where accretion is mediated solely by the MD's tenuous stellar wind, which has a mass-loss rate of $\dot{M}_c \lesssim 10^{-14}\,M_\odot\,\text{yr}^{-1}$ \citep{Wood05}, resulting in a relatively clean magnetospheric environment. Within this regime, if the WD's magnetic field dominates at the location of its companion ($\mu_s/a^3 \gg B_c$), the unipolar-inductor mechanism is activated (see Section~\ref{stage2} in detail). Therefore, the primary condition for activating the unipolar-inductor mechanism can be expressed as
\begin{align} 
a&\lesssim a_{\rm UI}\equiv\fraction{\mu_s}{B_c}{1/3}\simeq4.6\times 10^{10}~{\rm cm}
\nonumber\\ 
&\times\fraction{\mu_s}{10^{34}~{\rm G~cm^3}}{1/3}\fraction{B_c}{10^2~{\rm G}}{-1/3}.\label{acr}
\end{align} 
The characteristic surface magnetic fields for the binary components are $B_s \sim (10^6-10^9)$\,G for the magnetic WD and $B_c \sim (10^2-10^4)$\,G for the MD \citep[e.g.,][]{Ferrario15, Reiners22}. By combining Eq.(\ref{Kepler3}) and Eq.(\ref{acr}), we obtain
\begin{align} 
P&\lesssim P_{\rm UI}\equiv\left[\frac{4\pi^2 \mu_s}{G(M_s+M_c)B_c}\right]^{1/2}\simeq91~{\rm min}\nonumber\\ 
&\times\fraction{\mu_s}{10^{34}~{\rm G~cm^3}}{1/2}\fraction{M_s+M_c}{M_\odot}{-1/2}\fraction{B_c}{10^2~{\rm G}}{-1/2}.\label{Pui}
\end{align} 
If LPTs originate from the unipolar-inductor phase as discussed in Section \ref{stage2}, the observed 2--3 hour periods require that the WD should be magnetic.

3) Magnetosphere-interaction phase: This phase occurs at larger orbital separations, where the WD’s magnetic field at the location of the MD is significantly weaker. When this external field from the WD is much weaker than the MD's intrinsic surface field (i.e., $\mu_s/a^3 \ll B_c$), it can no longer penetrate the MD companion's surface. Instead, the two magnetospheres are forced to interact directly, leading to magnetic reconnection events that accelerate particles. This scenario is thus defined by the condition of
\begin{align}
a&\gtrsim a_{\rm UI}\simeq4.6\times 10^{10}~{\rm cm}
\nonumber\\ 
&\times\fraction{\mu_s}{10^{34}~{\rm G~cm^3}}{1/3}\fraction{B_c}{10^2~{\rm G}}{-1/3},
\end{align}
and the corresponding period constraint is
\begin{align} 
P&\gtrsim P_{\rm UI}\simeq91~{\rm min}\nonumber\\ 
&\times\fraction{\mu_s}{10^{34}~{\rm G~cm^3}}{1/2}\fraction{M_s+M_c}{M_\odot}{-1/2}\fraction{B_c}{10^2~{\rm G}}{-1/2}.\label{Pmi}
\end{align} 
In the following subsections, we evaluate each of these three scenarios, assessing their viability as LPT origins and deriving the associated physical constraints.

\subsection{Accretion phase}\label{stage1}

\subsubsection{Real accretion versus propeller process}

In the accretion phase, which occurs when the MD fills its Roche lobe ($a \lesssim R_c/\eta$), matter flows towards the WD, forming an accretion disk or accretion stream (panel (a) of Figure~\ref{threemodel}). The WD's magnetic field governs the inner region of this flow, dominating where its magnetic energy density becomes comparable to the kinetic energy density of the in-falling material. This critical boundary is known as the Alfv\'{e}n radius, $r_A$, and is given by
\be 
\frac{B^2(r_A)}{8\pi}\simeq\frac{1}{2}\rho(r_A)v^2(r_A). 
\ee 
We consider a steady transonic flow at nearly free-fall velocity, i.e., $v(r)=(2GM_s/r)^{1/2}$ and $\rho(r)=\dot M/4\pi r^2v$, where $\dot M$ is the mass loss rate of the MD companion. Then the Alfv\'{e}n radius is 
\begin{align}
r_A&=\fraction{\mu_s^4}{2GM_s\dot M^2}{1/7}=1.4\times10^{11}~{\rm cm}\fraction{\mu_s}{10^{34}~{\rm G~cm^3}}{4/7}\nonumber\\
&\times\fraction{M_s}{0.8M_\odot}{-1/7}\fraction{\dot M}{\dot 10^{-10}M_\odot~{\rm yr^{-1}}}{-2/7}.
\end{align}
In WD--MD binaries undergoing Roche-lobe overflow, the mass transfer rate, $\dot{M}$, is primarily determined by the mechanism of orbital angular momentum loss. For systems where magnetic braking dominates, the rate is typically $\dot M\sim(10^{-9}-10^{-8})M_\odot~{\rm yr^{-1}}$. For systems where gravitational radiation dominates, the rate is significantly lower, at $\dot M\sim(10^{-11}-10^{-10})M_\odot~{\rm yr^{-1}}$ \citep{Knigge11}.
The morphology of the accretion flow depends on the extent of the WD's magnetosphere relative to the binary separation. If the Alfv\'{e}n radius is large enough to intercept the accretion stream near the L1 point, $r_A \gtrsim a - R_{\rm L1} = (1-\eta)a$, the plasma will be channeled directly along the WD's magnetic field lines from the outset. This prevents the formation of an accretion disk and results in a stream accretion geometry, characteristic of Polar systems (see panel (a.1) of Figure \ref{threemodel}), and the corresponding condition is therefore
\begin{align} 
r_A=\fraction{\mu_s^4}{2GM_s\dot M^2}{1/7}
\gtrsim (1-\eta)a.
\end{align} 
Alternatively, if the Alfv\'{e}n radius is smaller than the distance to the L1 point ($r_A \lesssim (1-\eta)a$), the inflowing material possesses enough angular momentum to form an accretion disk. However, the WD's magnetic field will still dominate the innermost region, truncating the disk at $r \sim r_A$ (see panel (a.2) of Figure~\ref{threemodel}). This magnetically-gated disk accretion is the defining characteristic of IP systems.
The existence of both stream- (Polar) and disk- (IP) accretion geometries is consistent with the observed diversity within the mCV population. Observationally, Polars are predominantly found at shorter orbital periods ($P \lesssim (3-4)$ hrs), while IPs tend to prevail at longer periods (see Figure~\ref{CVperiod}). The fundamental physical distinction between these two subclasses remains a topic of debate: it is unclear whether IPs simply represent a population with intrinsically weaker magnetic fields than Polars, or if they possess comparable field strengths but have not yet achieved the spin-orbit synchronization characteristic of Polars.

The infalling plasma couples to the magnetic field of the WD. The outcome of this interaction depends on whether the WD's magnetospheric boundary, $\min[r_A, (1-\eta)a]$, lies inside or outside the co-rotation radius of the spinning WD, $r_{\rm co}$. The co-rotation radius is defined by
\begin{align}
r_{\rm co} &= \left(\frac{GM_s}{\Omega_s^2}\right)^{1/3}\simeq4.6\times10^{10}~{\rm cm}  \nonumber \\
&\times\left(\frac{M_s}{0.8\,M_\odot}\right)^{1/3} \left(\frac{P_s}{100\,\text{min}}\right)^{2/3}.
\end{align}
If the magnetospheric boundary is outside the co-rotation radius ($\min[r_A,(1-\eta)a] > r_{\rm co}$), the centrifugal force acting on the incoming plasma exceeds the gravitational force. This results in the accreting material being expelled from the system in a process known as the ``propeller effect''. In this scenario, although the MD fills its Roche lobe, the plasma does not penetrate the WD's magnetosphere, which may consequently remain relatively clean.
Conversely, if $\min[r_A,(1-\eta)a] < r_{\rm co}$, the system undergoes what can be termed ``real accretion''. The inflowing material from the Roche-lobe overflow fills the WD's magnetosphere, typically forming a truncated accretion disk. 
As will be discussed, such a process might suppress the formation of non-thermal (with population inversion) particle distribution responsible for the coherent radiation.

\subsubsection{Can Coherent Radiation be produced in an accretion environment?}

Several lines of evidence suggest that LPTs are unlikely to originate from systems in a state of stable accretion, i.e., where $\min[r_A,(1-\eta)a] < r_{\rm co}$. The primary argument stems from the nature of the emission itself. 
The generation of intense, coherent radio waves, as observed in LPTs, necessitates a relatively clean, low-density plasma environment.
The particles in a high-density accretion flow are significantly thermalized, thereby suppressing any coherent radiation mechanism. 
In this section, we discuss the critical condition under which the induced electric fields can substantially modify the particles' thermal distribution, a prerequisite for producing the population inversion that gives rise to maser radiation.
We consider there is a relative angular velocity, $\delta\Omega_r$, between the accreting plasma and the WD's magnetic field at a given radius $r$. The induced electric field at the plasma rest frame can be estimated as\footnote{Consider that the plasma has a non-relativistic velocity $\bm{v}$ in a background magnetic field $\bm{B}$. In the plasma rest frame, the electric field is $\bm{E}'\simeq\bm{v}\times\bm{B}/c$ and the magnetic field is $\bm{B}'\simeq\bm{B}$.}
\begin{align} 
E_{{\rm ind}}'&\sim\frac{\delta\Omega_r B(r)}{c}r\simeq\frac{\mu_s\delta\Omega_r}{r^2c}\nonumber\\ 
&\simeq0.14~{\rm statV~cm^{-1}}\fractionz{\mu_s}{10^{34}~{\rm G~cm^3}}\fraction{P}{100~{\rm min}}{-7/3}\nonumber\\ 
&\times\fraction{M_s+M_c}{M_\odot}{-2/3}\fraction{r}{a}{-2}\fractionz{\delta\Omega_r}{\Omega}.\label{ind}
\end{align}
where $\delta\Omega_r$ and $r$ are normalized to the orbital angular velocity $\Omega$ and semi-major axis $a$, respectively. 
For relatively weak $E_{\rm ind}'$, although its presence can induce an electromotive force or even currents, the particles' distribution in the plasma remains largely thermalized due to collisional processes, approximately following a Maxwellian distribution. However, a coherent radiation mechanism requires that the particle distribution is non-thermal and with population inversion. Therefore, we need to determine the critical field above which $E_{\rm ind}'$ significantly alters the particles' thermal distribution. 
In an unmagnetized plasma, this critical field, $E_{{\rm cr}}'$, might be characterized by
\be 
E_{{\rm cr}}'\sim\frac{k_{B} T}{e \lambda_{D}},\label{allow0} 
\ee 
where $\lambda_D = (k_BT / 4\pi e^2 n_e)^{1/2}$ is the Debye length, $T$ is the plasma temperature, and $n_e$ is the electron density within the accretion flow at a radius $r$ from the WD,
\begin{align}
n_e(r)&\simeq\frac{\dot M}{4\pi\mu_mm_pvr^2}\simeq1.6\times10^{9}~{\rm cm^{-3}}\fraction{P}{100~{\rm min}}{-1}\nonumber\\ 
&\times\fraction{M_s+M_c}{M_\odot}{-1/2}\fractionz{\dot M}{10^{-10}M_\odot~{\rm yr^{-1}}}\nonumber\\ 
&\times\fraction{M_s}{0.8M_\odot}{-1/2}\fraction{r}{a}{-3/2}.\label{edensity0}
\end{align}
Here, $\mu_m = 1.2$ is the mean molecular weight for solar composition, and $v \simeq (2GM_s/r)^{1/2}$ is the free-fall velocity of the inflowing material. 
Physically, $E_{\rm {cr}}'$ represents the theoretical limit beyond which the plasma can no longer maintain its quasi-neutrality and thermalization. This limit arises from the equilibrium between the plasma's internal thermal energy density and the energy density of an applied electrostatic field. When an induced field $E_{\rm ind}'$ approaches or exceeds this magnitude, the collective response of charged particles, driven by thermal motion, becomes insufficient to screen the field effectively. Consequently, the plasma's shielding capacity is overcome, leading to macroscopic charge separation and a breakdown of quasi-neutrality, thereby transitioning the particle distribution into a non-thermal regime\footnote{It should be noted that the field strength of $E_{\rm cr}'$ is much higher than that of the Dreicer field. When the electric field exceeds the Dreicer field, a small part of electrons in the high-energy tail can overcome collisional drag and undergo runaway acceleration. Although a fraction of high-energy electrons become non-thermal, the majority of electrons remain thermal. However, if the field exceeds $E_{\rm cr}'$, charged particles in the plasma would become separated, resulting in a breakdown of quasi-neutrality. In this case, a significant fraction of electrons will be accelerated and become non-thermal.} .
For the accreted plasma the critical electric field at $r$ is then
\begin{align} 
E_{{\rm cr}}'&\simeq0.16~{\rm statV~cm^{-1}}\fraction{T}{10^4~{\rm K}}{1/2}
\fraction{P}{100~{\rm min}}{-1/2}\nonumber\\ 
&\times\fraction{M_s+M_c}{M_\odot}{-1/4}\fraction{\dot M}{10^{-10}M_\odot~{\rm yr^{-1/2}}}{1/2}\nonumber\\ 
&\times\fraction{M_s}{0.8M_\odot}{-1/4}\fraction{r}{a}{-3/4}.
\label{allow}
\end{align} 
Although Eq.(\ref{allow0}) and Eq.(\ref{allow}) are estimated for an unmagnetized plasma, it also roughly applies to the magnetized accreted plasma we are interested here due to the following reasons:
In the plasma rest frame, the motion of particles is initially dominated by the induced electric field $E_{\rm ind}'$. As electrons are gradually accelerated, the Lorentz force becomes increasingly significant, resulting in an ``$\bm{E}\times\bm{B}$ drift'' with a velocity
\begin{align}
v_d &= \frac{E_{\rm ind}'}{B'}c \sim r \delta\Omega_r \simeq 520~{\rm km~s^{-1}} \nonumber\\
&\times\left(\frac{P}{100~{\rm min}}\right)^{-1/3} \left(\frac{M_s + M_c}{M_\odot}\right)^{1/3} \fractionz{\delta\Omega_r}{\Omega} \fractionz{r}{a}.
\end{align}
This drift velocity also represents, in order of magnitude, the enhanced speed of electrons due to the induced field. By comparison, the initial thermal velocity of electrons in the plasma rest frame is
\begin{align}
v_T \sim \fraction{3k_B T}{m_e}{1/2} \simeq 670~{\rm km~s^{-1}} \left(\frac{T}{10^4~{\rm K}}\right)^{1/2}.
\end{align}
Because $v_d \sim v_T$, although the magnetic field limits the acceleration of electrons in the plasma, it is evident that the energy distribution of electrons is significantly altered. Therefore, $E_{\rm ind}'>E_{\rm cr}'$ as the critical condition remains reasonable to a certain extent for an unmagnetized plasma.

In conclusion, the condition $E_{{\rm ind}}' \gtrsim E_{\rm{cr}}'$ serves as the criterion of the ability that the electric field can significantly change the thermal distribution of the electrons in the accreted plasma. This criterion can be translated into a critical accretion rate, $\dot{M}_{\rm cr}$, above which coherent radiation might be significantly quenched: 
\begin{align}
\dot M&\gtrsim\dot M_{\rm cr}\simeq7.6\times10^{-11}M_\odot~{\rm yr^{-1}}\fraction{\mu_s}{10^{34}~{\rm G~cm^3}}{2}\nonumber\\ 
&\times\fraction{P}{100~{\rm min}}{-11/3}\fraction{M_s+M_c}{M_\odot}{-5/6}\fraction{M_s}{0.8M_\odot}{1/2}\nonumber\\
&\times\fraction{T}{10^4~{\rm K}}{-1}\fraction{r}{a}{-5/2}\fraction{\delta\Omega_r}{\Omega}{2}.\label{dotMc}
\end{align} 
This condition is met across a vast parameter space for magnetic WD--MD binaries undergoing Roche-lobe overflow. Our analysis therefore suggests that the high-density environment associated with accretion rates of $\dot{M} \gtrsim 10^{-10}\,M_\odot\,\text{yr}^{-1}$ is fundamentally hostile to the generation of coherent radio emission. This theoretical conclusion is strongly supported by observational evidence from multiple fronts.
First, within the mCV population itself, intense coherent radio emission (brightness temperature $T_B \gtrsim 10^{12}$ K) is notably rare, rather than being a universal feature of Polars and IPs. Furthermore, an analogous suppression mechanism is observed in neutron star binaries. Studies of low-mass X-ray binaries (LMXBs) and their transition to millisecond pulsars reveal that the accretion phase is intrinsically unfavorable for producing coherent radio pulses \citep{Papitto13,Tauris18}). 

\begin{figure}
    \centering
    \includegraphics[width = 1.0\linewidth, trim = 40 100 10 50, clip]{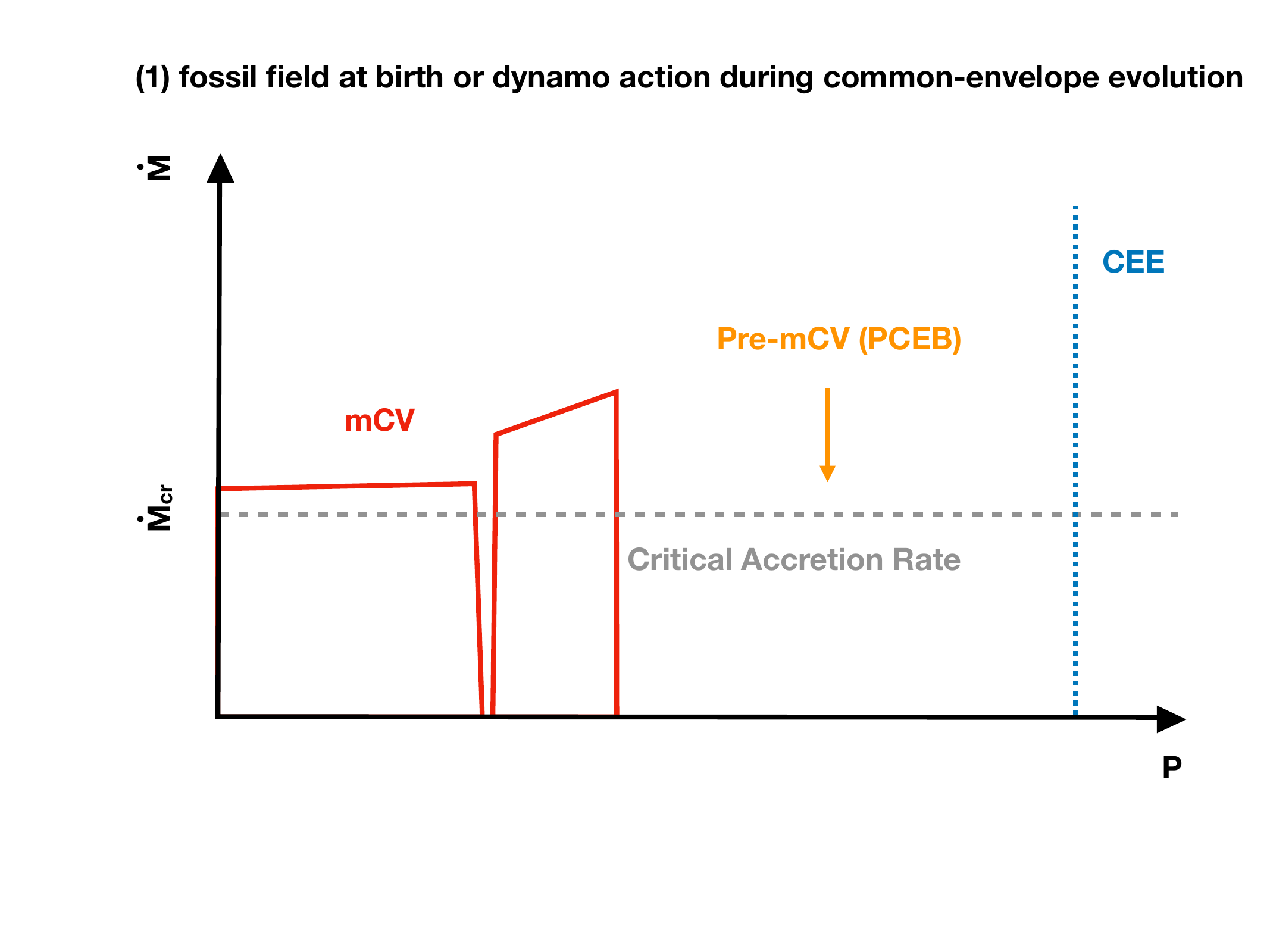}
    \includegraphics[width = 1.0\linewidth, trim = 40 100 10 50, clip]{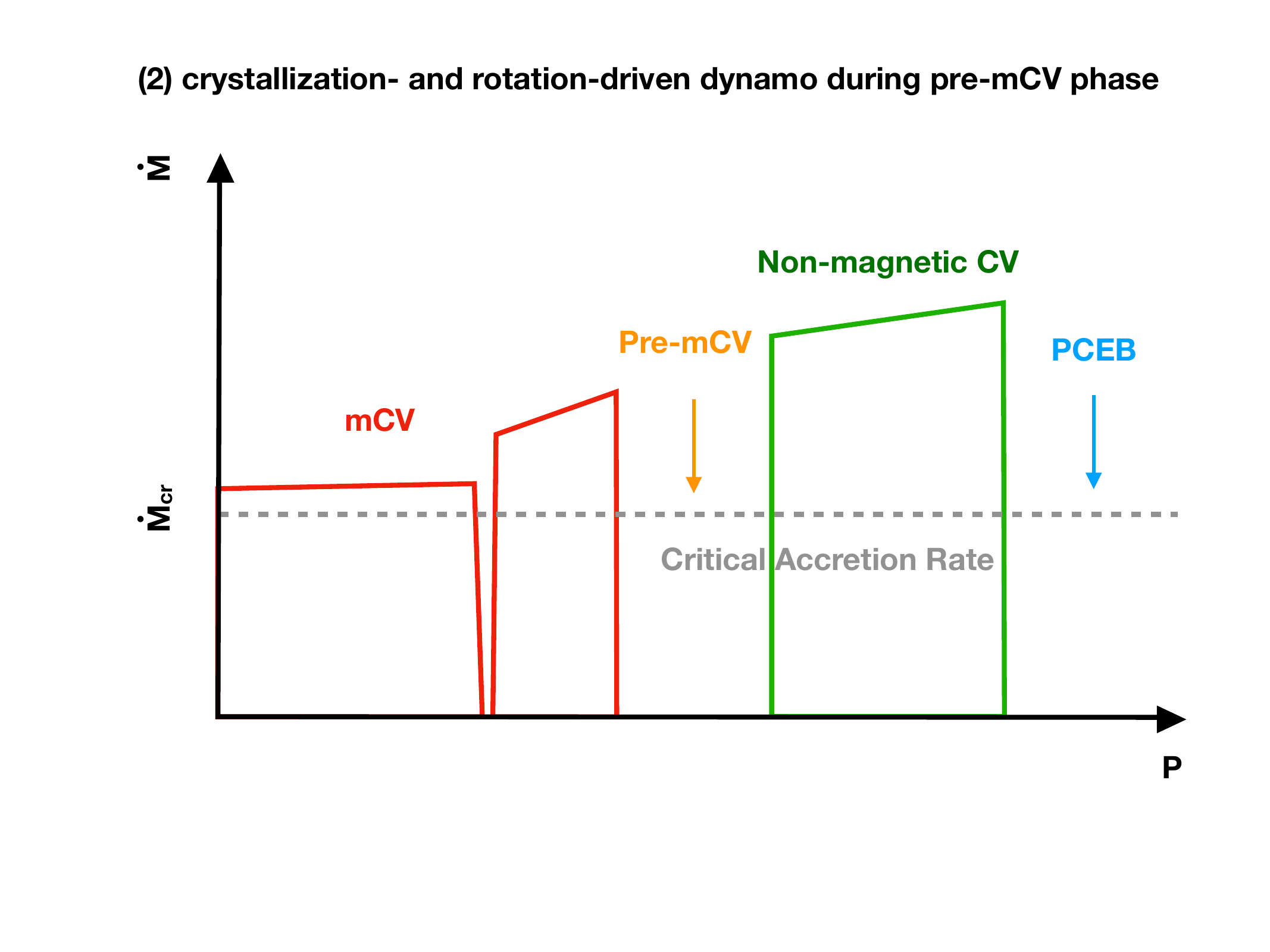}
    \caption{Schematic evolution of the mass transfer rate ($\dot{M}$) as a function of orbital period for a WD--MD binary. 
    Top panel: Evolution of the mass transfer rate assuming the WD magnetic field originates from the fossil field at birth \citep{Braithwaite04} or the dynamo action during common-envelope evolution (CEE) \citep{Tout08}. The pre-mCV phase includes the post-common-envelope binary (PCEB) phase.
    Bottom panel: Evolution of the mass transfer rate assuming the WD magnetic field originates from the crystallization- and rotation-driven dynamo, adapted from \citet{Schreiber21}.
    In this scenario, the WD acquires its magnetic field during the pre-mCV phase, marking the evolutionary transition from a non-magnetic CV to a fully magnetic CV.
    The evolutionary sequence proceeds from a PCEB, non-magnetic CV, pre-mCV to mCV. 
    In the pre-mCV phase of each possible scenario, the accretion rate can be significantly lower than the critical value $\dot{M}_{\rm cr}$ for typical parameters (Eq.(\ref{dotMc})), as indicated by the horizontal dashed line.}\label{accrate}
\end{figure} 

The preceding analysis suggests that stable accretion is hostile to coherent radio emission. Therefore, if such emission were to arise from a system undergoing Roche-lobe overflow, it would almost certainly have to be in the ``propeller'' regime with $\min[r_A,(1-\eta)a] > r_{\rm co}$. In this scenario, the centrifugal barrier expels the bulk of the accreting plasma, potentially leaving the inner magnetosphere ($r < \min[r_A,(1-\eta)a]$) sufficiently clean for coherent radiation mechanisms to operate. The prototypical examples of such a system are AE Aquarii (AE Aqr) and LAMOST J024048.51+195226.9, the only two confirmed IPs known to be in the propeller state \citep{Wynn97,Garnavich21,Pelisoli22}. For example, AE Aqr exhibits radio flares with a brightness temperature of $T_B \sim 10^{11} (r/10^{11}\,\text{cm})^{-2}$ K. For this emission to be coherent ($T_B > 10^{12}$ K), the emission region would need to be confined to $r < 10^{10}$ cm \citep{Bastian88}. However, key signatures of coherent radiation, such as rapid temporal variability or a high degree of circular polarization, are conspicuously absent in all observations of AE Aqr.
Furthermore, even if coherent radiation could be generated in the propeller phase, this scenario presents a fundamental conflict with the observed properties of LPTs. In a propeller-driven model, the emission would naturally be modulated by the WD's rotation, implying that the observed LPT period should correspond to the WD's spin period ($P_{\rm obs} \simeq P_s$). The energy source for the emission would likewise be the WD's rotational spin-down power of 
\begin{align}
L_{\rm sd}&\simeq\frac{\mu_s^2\Omega^4}{6c^3}\simeq7.4\times10^{23}~{\rm erg~s^{-1}}\nonumber\\ 
&\times\fraction{\mu_s}{10^{34}~{\rm G~cm^3}}{2}\fraction{P}{100~{\rm min}}{-4}.
\end{align}
This estimated spin-down power is insufficient to account for the observed isotropic luminosities of LPTs. Even if a highly optimistic beaming factor of $\sim 10^{-4}$ is invoked to reduce the intrinsic luminosity requirement, the available spin-down power remains barely comparable to the radiated power. When a realistic radiation efficiency ($\eta_{\rm rad}\ll 1$) is further considered, the energy budget becomes untenable. Therefore, the propeller mechanism is also disfavored as a viable origin for LPTs on energetic grounds. 

In conclusion, the low accretion rate condition suggests that LPTs most likely originate from the pre-mCV phase of WD–MD binary evolution. The characteristics of the pre-mCV phase depend on the underlying hypothesis for the generation of the WD's magnetic field. 
If the field originates from the fossil field at birth \citep{Braithwaite04} or the dynamo action during common-envelope evolution \citep{Tout08}, the pre-mCV phase can persist for a relatively long duration, potentially encompassing the post-common-envelope binary (PCEB) stage, as shown in the top panel of Figure~\ref{accrate}. Meanwhile, substantial interaction between the WD's magnetic field and the MD companion or its fields occurs and triggers LPTs only once the orbit has shrunk sufficiently. During this phase, the accretion rate remains low (driven solely by the weak stellar wind from the underfilling companion) but the magnetic interaction is significant.
In contrast, if the field is generated by a crystallization- and rotation-driven dynamo during the evolution from a non-magnetic CV to a fully magnetic CV \citep{Schreiber21}, the pre-mCV phase is much shorter, as shown in the bottom panel of Figure~\ref{accrate}. It is initiated by the emergence of a strong magnetic field on the WD, which drives magnetospheric interactions that widen the binary separation and detach the MD companion from its Roche lobe. Consequently, the high accretion rates characteristic of the preceding CV phase cease, leading to a sharp drop in accretion. This phase continues until orbital angular momentum losses shrink the orbit enough to re-establish mass transfer, marking the onset of the mature mCV phase. Due to the transient nature of the pre-mCV phase in this scenario, only a fraction of binaries are expected to be observed in the pre-mCV state.

The observed properties of ILT J1101+5521 and GLEAM-X J0704-37 align well with the theoretical picture of the pre-mCV phase. Their relatively long orbital periods (2-3 hours) and the absence of strong accretion signatures strongly suggest that they are currently in this detached, low-accretion state, with a mass transfer rate constrained to $\dot{M} \lesssim 10^{-10}M_\odot{\rm yr^{-1}}$. This hypothesis can be further tested for self-consistency. According to Eq.(\ref{Pacc1}), if ILT J1101+5521 and GLEAM-X J0704-37 are not in the accretion phase, their MD companions should have masses of $M_c \lesssim 0.2\,M_\odot$ and $M_c \lesssim 0.3\,M_\odot$, respectively. These theoretically required masses are approximately in agreement with observational constraints for these systems.
At last, the WDs in both ILT J1101+5521 and GLEAM-X J0704–37 exhibit very low effective temperatures, $T_{\rm eff} \lesssim 7500~{\rm K}$ \citep{deRuiter25,Rodriguez25}, which also supports that they are not in the accretion phase, see Section \ref{WDcooling} in detail. 

\subsection{Unipolar-inductor phase}\label{stage2}

\begin{figure}
    \centering
    \includegraphics[width = 0.9\linewidth, trim = 50 100 50 100, clip]{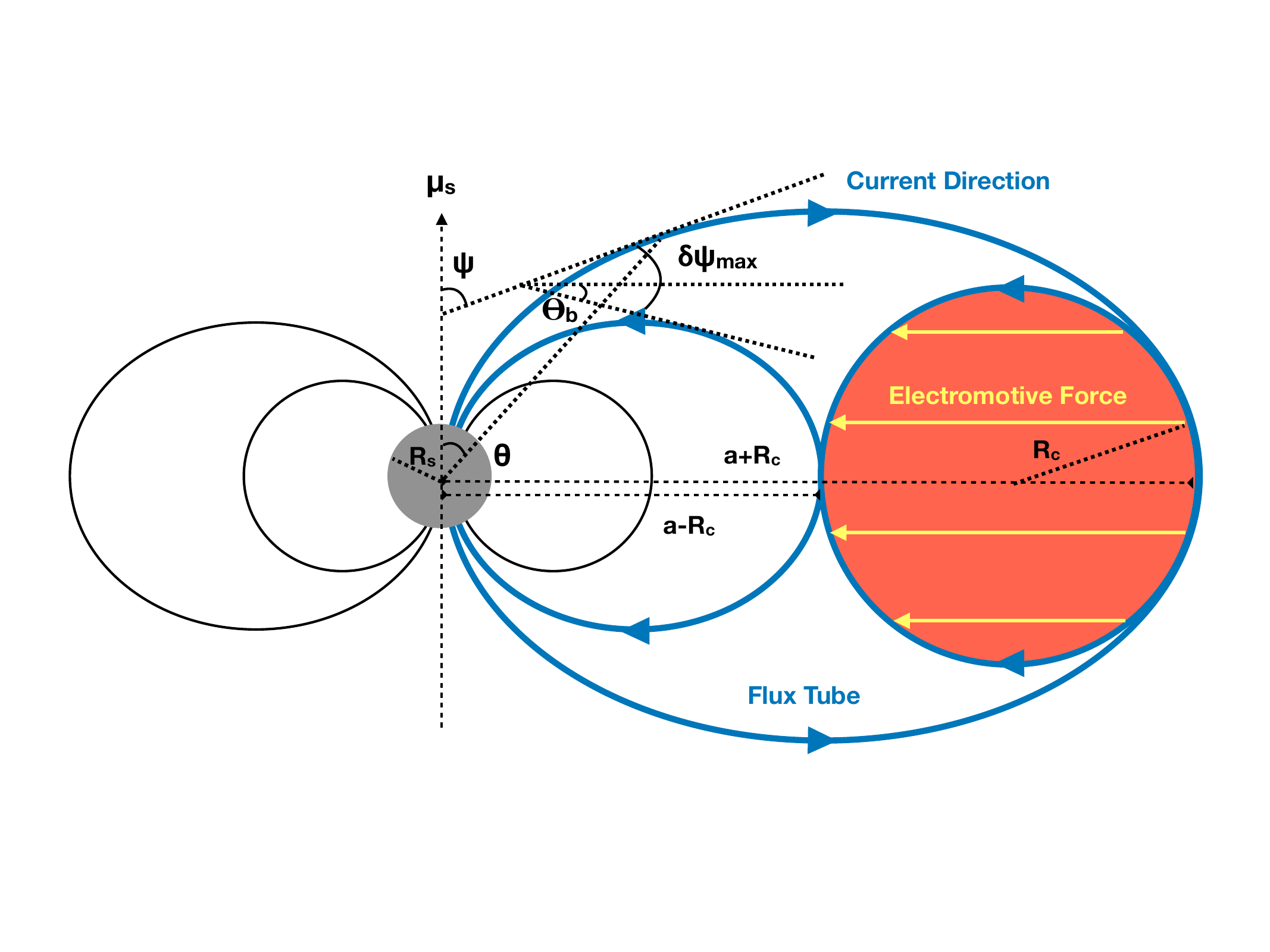} 
    \caption{Schematic diagram of the unipolar-inductor mechanism applied to a WD -- MD binary. The magnetic WD is assumed to have a dipole field, threading the non-magnetic MD. In the frame of the non-magnetic MD, there is an induced electric field of $\bm{E}'=(\bm{v}_{\rm rel}\times\bm{B})/c$. $\theta$ denotes the polar angle, $\psi$ denotes the angle between the field line and the magnetic moment $\bm\mu_s$, and $\delta\psi_{\max}$ denotes the angle difference of $\psi$ that can cover the MD companion.}\label{UI} 
\end{figure}

The unipolar-inductor mechanism operates in a specific regime where the binary is detached ($a \gtrsim R_c/\eta$) yet still compact enough for the WD's magnetic field to dominate the interaction ($a \lesssim a_{\rm UI}$). In this configuration, Roche-lobe overflow has ceased, and the WD's magnetosphere completely overwhelms the MD's intrinsic surface field ($B_c \ll \mu_s/a^3$). The orbital motion of the MD through the WD's powerful, rotating magnetosphere induces a substantial electromotive force (EMF). This EMF drives a current between the two stars, with the closed magnetic field lines serving as the conducting circuit (see Figure~\ref{UI} for a schematic).
This fundamental process of unipolar induction is not unique to WD--MD binaries; it has been successfully invoked to explain phenomena across a vast range of astrophysical scales. Well-known examples include the Jupiter-Io system \citep{Goldreich69}, ultracompact double WD binaries \citep{Wu02,DallOsso06}, the inspiral phase of binary neutron star or neutron star-black hole mergers \citep{Hansen01,McWilliams11,Lai12,Piro12,Wang16,Wang18}, and even close-in extrasolar super-Earths orbiting their host stars \citep{Laine12}.

A foundational requirement for the unipolar-inductor mechanism is asynchronism, i.e., a relative velocity between the WD's co-rotating magnetosphere and the orbital motion of the MD. This condition is fundamental to inducing an electromotive force (see Section~\ref{asynchronism} for a detailed discussion). Interestingly, this distinction between synchronous and asynchronous rotation defines the two primary subclasses of mCVs, i.e., Polars and IPs, respectively.
This dichotomy within the mCV population offers a crucial clue to the nature of LPTs. 
The magnetic fields of the IPs are at the low-field end of the distribution and partially overlap with the low field Polars, and IPs' orbital periods is $\gtrsim (3-4)$ hr, see Figure 9 of \citet{Ferrario15}. The asynchronism feature of IPs hints that LPTs from the WD -- MD binaries are the objects in the border regions of IPs but not in the accretion phase with the Roche-lobe overflow.

Combing Eq.(\ref{Pacc}) and Eq.(\ref{Pui}), a beautiful relation is obtained, 
\be 
\frac{P_{\rm UI}}{P_{\rm acc}}\simeq\fraction{\mu_s}{\mu_c}{1/2}\eta^{3/2}.
\ee 
If the LPTs can only be produced by the unipolar-inductor mechanism, $P_{\rm acc}<P<P_{\rm UI}$ is required, leading to the condition of
\be 
\frac{\mu_s}{\mu_c}>\eta^{-3}.\label{musmuc}
\ee 
The observed distributions of magnetic moments for WDs and MDs exhibit overlap. This overlap implies that a non-negligible fraction of WD--MD binaries should satisfy the condition of Eq.(\ref{musmuc}), making this a potentially important evolutionary channel. To quantify this fraction, we model the magnetic moments of both populations as log-normal distributions. Specifically, we assume that $\log \mu_s$ and $\log \mu_c$ follow normal distributions, $\mathcal{N}(\langle\log \mu_s\rangle, \sigma_{\log\mu_s})$ and $\mathcal{N}(\langle\log \mu_c\rangle, \sigma_{\log\mu_c})$, respectively.
Under this assumption, the variable $\log(\mu_s/\mu_c) = \log\mu_s - \log\mu_c$ also follows a normal distribution of $\log(\mu_s/\mu_c) \sim \mathcal{N}(\left<{\log \mu_s}\right>-\left<{\log \mu_c}\right>,(\sigma_{\log\mu_s}^2+\sigma_{\log\mu_c}^2)^{1/2})$.
The condition for unipolar-inductor mechanism given in Eq.(\ref{musmuc}) can be expressed logarithmically as $\log(\mu_s/\mu_c) > -3\log\eta\simeq(1.2-2.1)$. The fraction of WD--MD binaries satisfying this criterion can therefore be estimated by 
\be 
p_{\rm UI}=1-\Phi\left[\frac{-3\log\eta-(\left<{\log \mu_s}\right>-\left<{\log \mu_c}\right>)}{\sqrt{\sigma_{\log\mu_s}^2+\sigma_{\log\mu_c}^2}}\right],
\ee 
where $\Phi$ is the cumulative distribution function of the standard normal distribution.
To perform this calculation, we adopt empirically-derived distributions for the magnetic moments. For magnetic WDs, the distribution of magnetic moments is well-described by a log-normal function with a mean and standard deviation of $\langle\log (\mu_s/10^{33}\,\text{G}\,\text{cm}^3)\rangle = 0.7 \pm 0.3$ \citep{Wu91}. This corresponds to a normal distribution in logarithmic space, which we take as $\log (\mu_s/\text{G}\,\text{cm}^3) \sim \mathcal{N}(33.7, 0.3)$.
For MDs, the distribution is less constrained. However, given the lack of a strong correlation between spectral type and magnetic field strength \citep[see Figure 16 in][]{Reiners12}, we can approximate the distribution of the magnetic moment as being similar to that of the surface magnetic field. Guided by recent observations \citep{Wanderley24}, we therefore model the MD magnetic moments with a log-normal distribution, $\log (\mu_c/\text{G}\,\text{cm}^3) \sim \mathcal{N}(34, \sigma_{\log\mu_c})$, treating the standard deviation, $\sigma_{\log\mu_c}$, as a free parameter. Using these distributions and considering a plausible range of $\eta$ parameter, $\eta \simeq(0.2-0.4)$, we estimate the fraction of binaries in the unipolar-inductor phase, $p_{\rm UI}$. For a broad distribution with $\sigma_{\log\mu_c} = 1$, we find $p_{\rm UI} \simeq (0.01-0.08)$. For a narrower distribution with $\sigma_{\log\mu_c} = 0.5$, the fraction is significantly smaller, $p_{\rm UI} \simeq (2\times10^{-5}-5\times10^{-3})$.

Next, we turn to the energy dissipation process within the unipolar-inductor process. 
We consider that the external field from the WD is $B\sim\mu_s/a^3$, the relative velocity between the orbital motion and the magnetosphere rotation is $v_{\rm rel}=a\Delta\Omega\equiv(\Omega-\Omega_s)a$.
Due to the motion of the companion relative to the magnetic field of the magnetic WD, an EMF will be generated, $\varepsilon\sim 2E'R_c$ with the electromagnetic field of $\bm{E}'=\bm{v}_{\rm rel}\times\bm{B}/c$ in the MD frame, 
\begin{align} 
E'&\simeq\frac{\mu_s\Delta\Omega}{ca^2}\simeq0.14~{\rm statV~cm^{-1}}\fractionz{\mu_s}{10^{34}~{\rm G~cm^3}}\nonumber\\ 
&\times\fraction{P}{100~{\rm min}}{-7/3}\fraction{M_s+M_c}{M_\odot}{-2/3}\fractionz{\Delta\Omega}{\Omega}. 
\end{align} 
Thus, the EMF is given by
\begin{align}
&\varepsilon\sim\frac{2\mu_sR_c\Delta\Omega}{ca^2}\simeq2.8\times10^9~{\rm statV}\fractionz{\mu_s}{10^{34}~{\rm G~cm^3}}\nonumber\\
&\times\fractionz{R_c}{10^{10}~{\rm cm}}\fraction{P}{100~{\rm min}}{-7/3}\fraction{M_s+M_c}{M_\odot}{-2/3}\fractionz{\Delta\Omega}{\Omega}.
\end{align} 
This EMF will drive a current along the magnetic field lines in the magnetosphere, connecting the WD and the MD through magnetic flux tubes, see Figure \ref{UI}. 

The current in the circuit is $I=\varepsilon/\mathcal{R}$ with the total resistance of the circuit of 
\be 
\mathcal{R}=\mathcal{R}_s+\mathcal{R}_c+2\mathcal{R}_{\rm mag},\label{resis}
\ee  
where $\mathcal{R}_s$, $\mathcal{R}_c$, and $\mathcal{R}_{\rm mag}$ are the resistances of the WD, the MD, and the magnetosphere, respectively. 
We first estimate the resistances of the WD and the MD.
Generally, an object with a conductivity of $\sigma$, a thickness of $\Delta h$, and an area of $A$ has a resistance $\mathcal{R}\sim \Delta h/\sigma A$.
Thus, the ratio of their resistances is given by
\be 
\frac{\mathcal{R}_s}{\mathcal{R}_c}\sim\fractionz{\Delta h_s}{\Delta h_c}\fraction{\sigma_s}{\sigma_c}{-1}f^{-1}\fraction{R_s}{R_c}{-2},\label{fracresis}
\ee 
where $\sigma_s$ and $\sigma_c$ are the conductivity of the WD and the MD, respectively, $\Delta h_s$ and $\Delta h_c$ the thickness of the dissipative surface layers of the WD and the MD, respectively, and $f$ is the fractional effective area of the magnetic poles (hot spots) on the surface of the WD. 
Since the current is mainly on the surfaces of the WD and MD, their conductivity can be approximately estimated by the Spitzer formula \citep{Spitzer53,Wu02}
\be 
\sigma\sim\sigma_{\rm Spitzer} = \frac{2^{5/2}}{\pi^{3/2}} \frac{ \gamma\left( k_B T \right)^{3/2}}{m_e^{1/2} Z e^2 \ln \Lambda},
\ee 
where $T$ is the electron temperature in the atmosphere, $Z$ is the charge number per ion, and $\ln\Lambda$ is the Coulomb logarithm. The factor $\gamma$ depends on $Z$ and can vary between 0.6 (for $Z = 1$) and 1 (in the limit $Z\rightarrow\infty$). 
The thickness of the dissipative surface layers $\Delta h$ should be on the same order of magnitude as the star surface's skin depth $\delta$, thus one has 
\be
\Delta h\sim\delta\propto\sigma^{-1/2}.
\ee
Thus, the ratio of the WD's resistance to that of the MD can be expressed as
\be 
\frac{\mathcal{R}_s}{\mathcal{R}_c}\sim\fraction{T_s}{T_c}{-9/4}\fraction{\ln\Lambda_s}{\ln\Lambda_c}{3/2}f^{-1}\fraction{R_s}{R_c}{-2}.
\ee 
For the WD and MD in a LPT system \citep{deRuiter25,Rodriguez25}, one usually has $T_s/T_c\sim 2-10$ (typical Polars have a WD mass of $\lesssim0.6M_\odot$ and a WD effective temperatures of $\lesssim11000$ K according to \citet{Townsley09}), $\ln\Lambda_s/\ln\Lambda_c\sim$ a few, and $R_s/R_c\lesssim0.1$ and $f\lesssim0.1$ (because the current only passes through a small portion of the WD surface, see Figure \ref{UI}). Thus, the effective resistance of the WD is significantly larger than that of the MD, $\mathcal{R}_s\gg\mathcal{R}_c$. Using the geometry of the binary and dipole field, the WD resistance is given by \citep{Wu02}
\begin{align}
\mathcal{R}_s&\simeq\frac{\mathcal{J}}{2\sigma_sR_c}\fractionz{H}{\Delta d}\fraction{a}{R_s}{3/2}\nonumber\\
&\simeq 1.7\times10^{-21}~{\rm s~cm^{-1}}\fraction{\sigma_s}{10^{13}~{\rm s^{-1}}}{-1}\fractionz{P}{100~{\rm min}}\nonumber\\ 
&\times\fraction{M_s+M_c}{M_\odot}{1/2}\fraction{R_s}{10^9~{\rm cm}}{-3/2}\fraction{R_c}{10^{10}~{\rm cm}}{-1},\label{Rs}
\end{align} 
where $H$ is the atmospheric depth where electrical currents cross the magnetic field lines before returning to the MD, $\Delta d$ represents the thickness of the current layer's arc-shaped cross-section within the WD's atmosphere, the factor of $J$ is determined by the radii of the stars in relation to their orbital separation, which typically approximates unity.
Here, we use the assumption of $\mathcal{J}\sim H/\Delta d\sim1$ \citep{Li98,Wu02,Piro12} and take the conductivity as $\sigma_s\sim(10^{13}-10^{14})~{\rm s^{-1}}$ for a WD atmosphere with $T\sim(10^4-10^5)$ K.

The magnetospheric contribution to the total circuit resistance is poorly constrained, as it depends critically on the complex plasma environment within the binary system. However, a fundamental lower limit can be placed on the total resistance of the circuit. As demonstrated by \citet{Lai12}, an excessively large current flow would generate a strong toroidal magnetic field, severely twisting the magnetic flux tube that connects the two stars. This magnetic stress would ultimately lead to a disruption of the current circuit. The stability of the circuit therefore imposes a lower limit on the total resistance, $\mathcal{R}_{\rm tot, lim}$. This limit is derived from the condition that the magnetic pitch angle, $\zeta_\phi \sim 16 v_{\rm rel} / c^2 \mathcal{R}_{\rm tot}$, must remain small ($\zeta_\phi < 1$), implying
\begin{align} 
\mathcal{R}_{\rm tot}&>\mathcal{R}_{\rm tot,lim}=\frac{16a\Delta\Omega}{c^2}\simeq9.2\times10^{-13}~{\rm s~cm^{-1}}\nonumber\\
&\times\fraction{P}{100~{\rm min}}{-1/3}\fraction{M_s+M_c}{M_\odot}{1/3}\fractionz{\Delta\Omega}{\Omega}.
\end{align}
Furthermore, if the magnetosphere resistance is given by the impedance of free space \citep{Piro12,Lai12}, one would have $\mathcal{R}_{\rm mag}\simeq4\pi/c=4.2\times10^{-10}~{\rm s~cm^{-1}}$ and $\zeta_\phi\sim4v_{\rm rel}/\pi c$. The condition of $\zeta_\phi<1$ lead to a criterion of
\be 
\frac{\Delta\Omega}{\Omega}<\fractionz{\Delta\Omega}{\Omega}_{\rm cr}\simeq450\fraction{P}{100~{\rm min}}{1/3}\fraction{M_s+M_c}{M_\odot}{-1/3}.
\ee
For $\Delta\Omega/\Omega\ll(\Delta\Omega/\Omega)_{\rm cr}$, the unipolar-inductor is unsaturated due to $\zeta_\phi\ll1$ and the energy dissipation rate of the system, $\dot E\simeq2\varepsilon^2/\mathcal{R}$, is
\begin{align} 
\dot E&
\simeq\zeta_{\phi}\Delta\Omega\frac{\mu_s^2R_c^2}{2a^5}
\stackrel{\rm FP}{=}\frac{2\mu_s^2R_c^2\Delta\Omega^2}{\pi c a^4}\nonumber\\
&\simeq3.9\times10^{28}~{\rm erg~s^{-1}}\fraction{\mu_s}{10^{34}~{\rm G~cm^3}}{2}\fraction{R_c}{10^{10}~{\rm cm}}{2}\nonumber\\ 
&\times\fraction{P}{100~{\rm min}}{-14/3}\fraction{M_s+M_c}{M_\odot}{-4/3}\fraction{\Delta\Omega}{\Omega}{2},\label{diss_up}
\end{align}
where ``FP'' denotes the assumption of the impedance of free space for $\zeta_\phi<1$. The isotropic luminosity of LPTs can reach $L_{\rm iso}\sim10^{28}~{\rm erg~s^{-1}}$ \citep{deRuiter25}. Therefore, for the unipolar-inductor mechanism to power the observed LPT luminosities, considering a realistic radiation efficiency with $\eta_{\rm rad}\ll 1$, the binary system must be highly asynchronous ($\Delta\Omega/\Omega \gg 1$), as pointed by \citet{Qu25}.

Furthermore, the power output of the unipolar-inductor mechanism does not increase indefinitely with asynchronicity. Once $\Delta\Omega/\Omega$ exceeds the critical threshold of $(\Delta\Omega/\Omega)_{\rm cr}$, the circuit would break down due to the twisting flux tube, i.e., $\zeta_\phi\sim1$. In this ``saturated'' regime, the unipolar-inductor mechanism operates at its maximum possible dissipation rate, given by 
\begin{align} 
\dot E&
=\zeta_{\phi}\Delta\Omega\frac{\mu_s^2R_c^2}{2a^5}
\stackrel{\zeta_\phi\sim1}{=}\Delta\Omega\frac{\mu_s^2R_c^2}{2a^5}\nonumber\\
&\simeq1.8\times10^{31}~{\rm erg~s^{-1}}\fraction{\mu_s}{10^{34}~{\rm G~cm^3}}{2}\fraction{R_c}{10^{10}~{\rm cm}}{2}\nonumber\\ 
&\times\fraction{P}{100~{\rm min}}{-13/3}\fraction{M_s+M_c}{M_\odot}{-5/3}\fractionz{\Delta\Omega}{\Omega}.\label{diss_up2}
\end{align} 

In summary, the unipolar-inductor mechanism exhibits two distinct scaling behaviors for the energy dissipation rate, $\dot{E}$, depending on the degree of asynchronicity. In the unsaturated regime ($\Delta\Omega/\Omega \ll (\Delta\Omega/\Omega)_{\rm cr}$), the dissipation rate is a strong function of the orbital period, scaling as $\dot E\propto P^{-14/3}(\Delta\Omega/\Omega)^{2}\propto P^{-8/3}$ for $P_s\ll P$. Conversely, in the saturated regime ($\Delta\Omega/\Omega \gg (\Delta\Omega/\Omega)_{\rm cr}$), the scaling becomes even steeper, following $\dot E\propto P^{-13/3}(\Delta\Omega/\Omega)\propto P^{-10/3}$ for $P_s\ll P$. As we will demonstrate in Section~\ref{LF}, this period-dependent energy dissipation rate is the fundamental driver that shapes the luminosity function of the LPT population.
It is important to note that these calculations are performed under the standard assumption for mCVs, namely that the orbital period is longer than the WD's spin period ($P \gtrsim P_s$). The alternative scenario, $P < P_s$, where the companion orbits faster than the primary rotates, is generally considered to be dynamically unstable and is not explored further in this work. 

At last, according to Eq.(\ref{resis}), Eq.(\ref{fracresis}), Eq.(\ref{Rs}), Eq.(\ref{diss_up}) and Eq.(\ref{diss_up2}), the energy dissipation rate at the surfaces of the WD and MD is about
\be 
\dot E_i\simeq\frac{\mathcal{R}_i}{2\mathcal{R}_{\rm mag}}\dot E,
\ee 
where $i=s,c$ for the cases of the WD and the MD, respectively. Thus, the energy dissipation rate at the surfaces is much smaller than the energy dissipation rate of the magnetosphere.

\subsection{Magnetosphere-interaction phase}\label{stage3}

\begin{figure}
    \centering
    \includegraphics[width = 1.0\linewidth, trim = 00 100 00 100, clip]{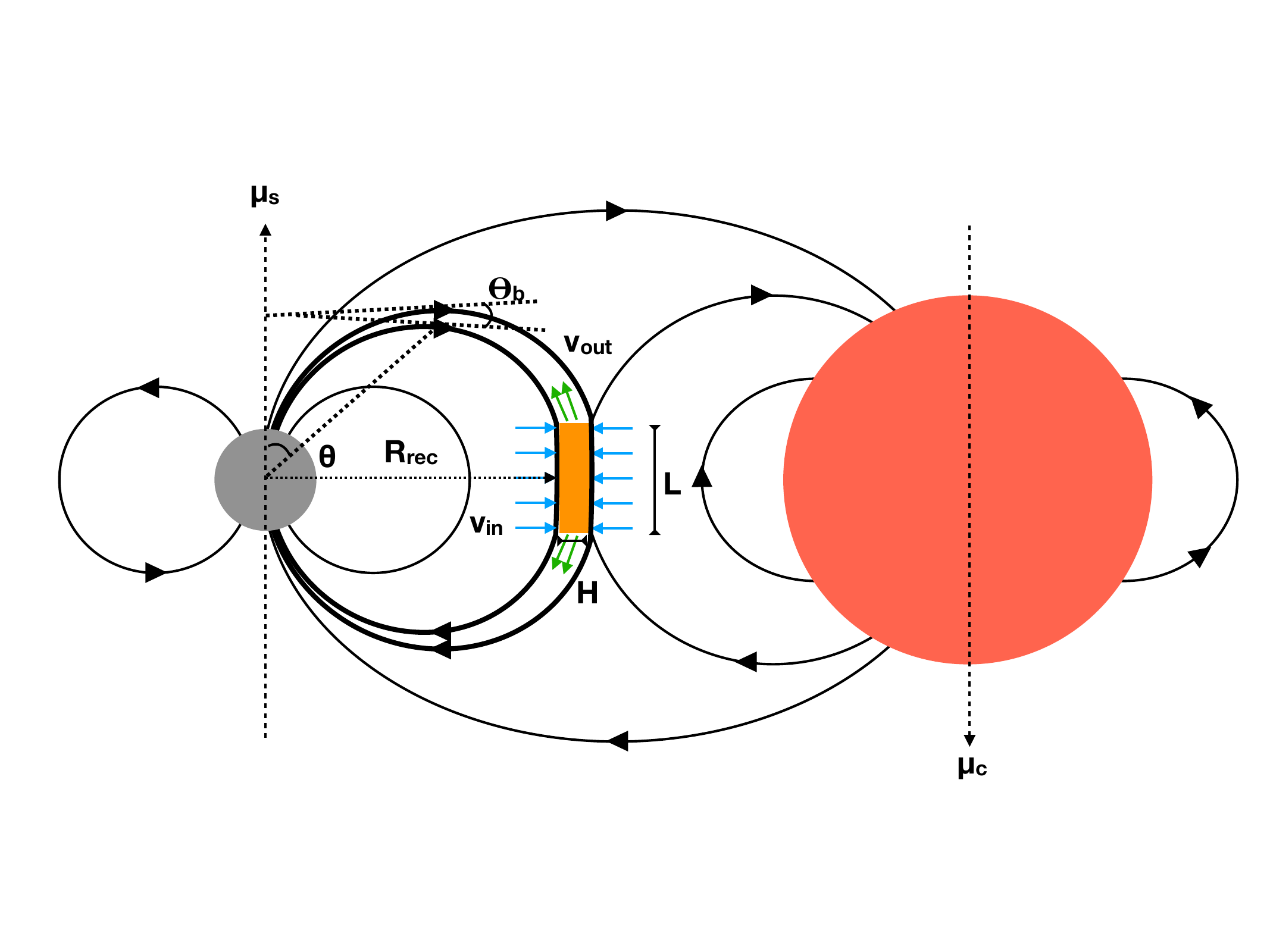} 
    \caption{Schematic diagram of the magnetosphere interaction applied to the WD -- MD binary. Both the magnetic WD and the magnetic MD are assumed to have a dipole field with magnetic moments of $\bm{\mu_s}$ and $\bm{\mu}_c$. The magnetic reconnetion region has a length scale of $L$, a thickness of $H$ and is at $R_{\rm rec}$ from the WD center. Particles are accelerated at the magnetic reconnection region and move to the stars along the field lines.}\label{MI} 
\end{figure}

In the magnetosphere-interaction phase with $a\gtrsim a_{\rm UI}$, the magnetic field of the MD surface is strong enough that the magnetic field from the WD cannot penetrate into its surface, $\mu_s/a^3\ll B_c$. In this scenario, the magnetic reconnection process dominates the energy dissipation, see Figure \ref{MI}. 
In astrophysics, the collisionless magnetic reconnection is the main fast magnetic reconnection mechanism, driven by kinetic-scale physics rather than by collisional resistivity. 
Instead of relying on particle collisions (resistivity), it's enabled by plasma kinetic effects operating at small ion and electron scales, such as the Hall effect and electron inertia or pressure.
Crucially, this allows reconnection to occur much faster than predicted by resistive models, efficiently converting stored magnetic energy into plasma heating, bulk kinetic energy, and energetic particle acceleration, thus explaining many explosive phenomena like solar flares and magnetospheric substorms.
The collisionless effects are important when the system scale is less than the critical length scale \citep{Zweibel09}
\begin{align}
L&<L_c\equiv4\times10^{12}~{\rm cm}~B\fraction{T}{n_e}{3/2}\simeq1.3\times10^{15}~{\rm cm}\nonumber\\ 
&\times\fractionz{\mu_s}{10^{34}~{\rm G}}\fraction{r}{10^{10}~{\rm cm}}{-3}\fraction{T}{10^4~{\rm K}}{3/2} \fraction{n_e}{10^5~{\rm cm}}{-3/2}.\label{LC}
\end{align} 
For the WD -- MD binary in the pre-mCV phase without the Roche-lobe overflow accretion, according to Eq.(\ref{edensity0}), the electron number density at $r$ in the magnetosphere can be estimated by 
\begin{align}
n_e(r)&\simeq1.6\times10^{5}~{\rm cm^{-3}}\fraction{P}{100~{\rm min}}{-1}\fraction{M_s+M_c}{M_\odot}{-1/2}\nonumber\\ 
&\times\fractionz{\dot M}{10^{-14}M_\odot~{\rm yr^{-1}}}\fraction{M_s}{0.8M_\odot}{-1/2}\fraction{r}{a}{-3/2}.
\label{edensity}
\end{align}
The MD's stellar wind, with a mass-loss rate of $\dot{M} \lesssim 10^{-14}\,M_\odot\,\text{yr}^{-1}$ \citep{Wood05}, is extremely tenuous. For typical system parameters, this corresponds to an electron density of $n_e \lesssim 10^5\,\text{cm}^{-3}$ at $r\sim a$, which ensures that the characteristic system size (also the characteristic length scale of the reconnection region), $L$, is much smaller than the above critical length scale, $L_c$ ($L \ll L_c$). This confirms that collisionless magnetic reconnection is indeed the dominant energy dissipation mechanism during the large-scale magnetospheric interaction in a WD--MD binary.
We assume that reconnection primarily occurs at a radius $R_{\rm rec}$ from the WD's center (see Figure~\ref{MI}), defined as the point where the magnetic fields of the WD and the MD are comparable in strength but differ significantly in direction, $\mu_s/R_{\rm rec}^3 \simeq \mu_c/(a-R_{\rm rec})^3$. Solving for $R_{\rm rec}$ yields
\be 
L\lesssim R_{\rm rec}\simeq a\left[\fraction{\mu_c}{\mu_s}{1/3}+1\right]^{-1}~~~{\rm for}~R_s<R_{\rm rec}<a.\nonumber\\\label{Rrec}
\ee 
In the majority of plausible scenarios for mCV systems, the magnetic moments of the two stars are expected to be of a similar order of magnitude (as discussed in Section~\ref{stage2}), leading to $(\mu_c/\mu_s)^{1/3} \sim 1$. Substituting this into the preceding equation, the reconnection radius, $R_{\rm rec}$, is comparable to the binary separation, $a$. This implies that magnetic reconnection is a global phenomenon, occurring on a scale comparable to the entire binary system.

The strong, stable periodicity observed in LPTs strongly suggests that their emission is controlled by a persistent, ordered magnetic geometry, rather than being shaped by a turbulent plasma environment.
This observational constraint directly implies that the emission region must be magnetically dominated, a condition quantified by the plasma $\beta$ parameter (i.e., the ratio of plasma gas pressure to magnetic pressure) being much less than unity ($\beta \ll 1$).
A low-$\beta$ environment is crucial for several reasons. First, the magnetic field structure remains stable and is not easily disrupted by plasma pressure gradients, providing the persistent geometric framework required for periodic emission. Second, it facilitates the formation of a well-defined X-point magnetic reconnection geometry. Third, the high magnetic energy density provides a substantial energy reservoir available for release. Fourth, reconnection-driven outflows can more effectively expel plasma, maintaining a low-density environment conducive to coherent radiation.
Finally, a low-$\beta$ environment is a prerequisite for triggering fast, collisionless magnetic reconnection. The macroscopic plasma dynamics in such a regime can efficiently compress current sheets down to kinetic scales (e.g., the ion inertia length), unleashing the most rapid and efficient form of magnetic energy conversion known. Given the strong magnetic fields characteristic of WDs and MDs, the low-$\beta$ condition is readily satisfied in these binary systems. This establishes magnetic reconnection as a highly plausible and effective energy dissipation mechanism for powering LPTs.

The differential rotation between the two magnetospheres dynamically modulates the magnetic reconnection region; that is, the reconnection region continuously evolves in response to this differential rotation. As a result, the observed energy dissipation rate varies with the beat period, see Section \ref{asynchronism}. Such a macroscopically modulated reconnection process differs from the transient, explosive magnetic reconnection associated with stellar flares or magnetar bursts. In the latter scenario, reconnection is typically triggered by instantaneous perturbations in the magnetosphere, evolving through plasmoid ejection and current sheet stretching \citep[e.g.,][]{Priest02}. In contrast, reconnection driven by magnetospheric interactions we are interested here is continuously present, with its location dynamically shifting over time.
In observations, both the pronounced periodicity and the small pulse duty cycles detected in LPTs suggest a geometric origin of the pulse profile, analogous to that of radio pulsars (see Section \ref{beaming}), which support the interpretation of a macroscopically modulated reconnection process.
Nonetheless, the pulse profile in each period may behave differently due to some instability processes, see the discussion at the end of this section.

The interaction between two stars' magnetospheres can be analyzed from two complementary perspectives: (i) a macroscopic viewpoint, which considers the global interaction between the two magnetic dipoles, and (ii) a microscopic viewpoint, which focuses on the kinetic plasma physics of the local magnetic reconnection process. 
The macroscopic approach can provide a clear and direct estimate of the total available power supply.
For simplicity, we consider only the interaction between the intrinsic dipole fields of the WD and the MD, neglecting any induced magnetic moments. Within this framework, the interaction exerts a synchronizing torque on the binary system, given by
\be 
N_B\sim\frac{\mu_s\mu_c}{a^3}.
\ee
Asynchronism ($\Delta\Omega \neq 0$) is a fundamental prerequisite for magnetic reconnection in this context. It is this relative motion between the two magnetospheres that drives plasma into the reconnection region, continuously supplying the energy needed for dissipation. Thus, the rate at which magnetic energy is converted into plasma energy through this process is estimated as
\begin{align} 
\dot E&=N_B\Delta\Omega\sim\frac{\mu_s\mu_c\Omega^2\Delta\Omega}{G(M_s+M_c)}\simeq8.6\times10^{31}~{\rm erg~s^{-1}}\nonumber\\
&\times\fractionz{\mu_s}{10^{34}~{\rm G~cm^3}}\fractionz{\mu_c}{10^{33}~{\rm G~cm^3}}\fraction{P}{100~{\rm min}}{-3}\nonumber\\ 
&\times\fraction{M_s+M_c}{M_\odot}{-1}\fractionz{\Delta\Omega}{\Omega}.\label{diss_mc}
\end{align} 
Due to the presence of the MD's magnetic field, the energy dissipation rate in the magnetosphere-interaction phase can, for a given $\Delta\Omega/\Omega$, be larger than that of the unsaturated unipolar-inductor process and comparable to that of the saturated process (discussed in Section~\ref{stage2}). Such a scenario therefore permits a lower radiation efficiency to explain the observed luminosities compared to what is required by the unipolar-inductor mechanism at a moderate $\Delta\Omega/\Omega$.

Based on its high efficiency and its applicability on the relevant physical scales (as confirmed by Eq.(\ref{LC})), we have identified collisionless magnetic reconnection as the primary particle acceleration mechanism during the magnetosphere-interaction phase. A key feature of this process, distinguishing it from resistive models, is that its reconnection rate is largely independent of the system's plasma resistivity \citep[e.g.,][]{Shay99, Liu17},
\be 
M_A\equiv \frac{v_{\rm in}}{v_{A,{\rm in}}}\sim(0.01-0.1),
\ee  
where $v_{\rm in}$ is the velocity of the plasma flowing into the reconnection zone, $v_{A,{\rm in}}=B_{\rm in}/(4\pi\rho_{\rm in})^{1/2}$ is the Alfv\'{e}n velocity in the inflow zone of the magnetic reconnection, and $\rho_{\rm in}$ is the plasma density in the inflow zone. The outflow velocity is 
\be 
v_{\rm out}\sim v_{A,{\rm in}},
\ee 
suggesting a high efficiency of converting magnetic energy into particle kinetic energy. 
The magnetosphere-interaction phase is characterized by a high magnetic field strength and a low plasma density. These conditions result in a relativistically high Alfv\'{e}n velocity for the typical parameters of these systems. 

Finally, we estimate the typical timescale of magnetic reconnection instability, which can be characterized by the Alfv\'{e}n timescale,
\begin{align} 
\tau_{\rm ins}&\sim \frac{L}{v_{A,{\rm in}}M_A}\sim\frac{L}{cM_A}\simeq (16-160)~{\rm s}\fraction{P}{100~{\rm min}}{2/3} \nonumber\\
&\times\fraction{M_s+M_c}{M_\odot}{1/3}\left[\fraction{\mu_c}{\mu_s}{1/3}+1\right]^{-1}.
\end{align}
Such a timescale approximately corresponds to the timescale of the structure of LPTs in the picture of the magnetosphere interaction.

\section{Radiation process}\label{radiation}

\subsection{Geometry on the emission region}\label{beaming}

The observed properties of LPTs, including their high brightness temperatures, periodic pulses, and significant linear and circular polarization, provide crucial diagnostics of the magnetic field geometry and physical conditions of the emission region. First, we can estimate the brightness temperature directly from typical observational parameters. For a canonical LPT, we adopt a peak flux density of $F_\nu \simeq 0.1$ Jy, a pulse duration of $\Delta t \simeq 100$ s, an observing frequency of $\nu \simeq 0.1$ GHz, and a distance of $d \simeq 0.1$ kpc.
Using these values, the brightness temperature, $T_B$, can be constrained by
\begin{align} 
T_{B}&\gtrsim T_{B,{\rm lim}} \equiv \frac{1}{2\pi k_B}\left(\frac{d}{\nu\Delta t}\right)^2 F_\nu\simeq10^{12}~{\rm K}\fraction{d}{0.1~{\rm kpc}}{2}\nonumber\\ 
&\times\fraction{\nu}{0.1~{\rm GHz}}{-2}\fraction{\Delta t}{100~{\rm s}}{-2}\fractionz{F_\nu}{0.1~{\rm Jy}}.\label{brightness}
\end{align} 
This estimation for $T_{B,{\rm lim}}$ implicitly assumes that the transverse size of the emission region, $l_{\rm em}$, is approximately equal to the light-travel distance over the pulse duration, thus, $l_{\rm em} \sim c\Delta t$. However, $c\Delta t$ sets only an upper limit on the true size of the source. The observed pulse duration, $\Delta t$, could be determined by the intrinsic timescale of the emission process itself, or by the transit time of a rotating beam across the line of sight (i.e., ``lighthouse effect'' in pulsars), both of which could be longer than the light-crossing time of the actual source, $\sim l_{\rm em}/c$. Therefore, the intrinsic transverse scale is likely smaller, $l_{\rm em} \lesssim c\Delta t$, which in turn implies that our calculated $T_{B,{\rm lim}}$ represents a robust lower limit on the actual brightness temperature. This distinction is critical when comparing this observational estimate to the intrinsic brightness temperature predicted by theoretical radiation mechanisms.
Furthermore, the standard brightness temperature formula inherently assumes a non-relativistic bulk motion for the emission region. This assumption is consistent with the scenario of an emission region co-rotating with the WD's magnetosphere, which is itself non-relativistic\footnote{It is important to distinguish the bulk motion of the emission region, which is non-relativistic here, from the intrinsic motion of the individual radiating particles, which can be highly relativistic. The standard brightness temperature calculation mainly requires correction for the former.}. 

Several observational properties of LPTs point towards a magnetically dominated emission environment. First, their high brightness temperatures imply a correspondingly high radiation energy density within the emission region. If this radiation is powered by magnetic dissipation, i.e., the unipolar-inductor or magnetosphere-interaction mechanisms, the magnetic energy density must exceed the radiation energy density to confine the plasma. Second, the strong periodicity and the small pulse duty cycles (which imply a beamed emission) suggest that the radiation pattern is rigidly controlled by a stable magnetic field geometry.
Let us assume the emission occurs at a radius $r_{\rm em}$ from the WD's center, where the ambient magnetic field is approximately $B_{\rm em}=\mu_s/r_{\rm em}^3$. The energy density of the electromagnetic wave of a LPT at $r_{\rm em}$ is $U_{\rm EM}=L_{\rm iso}/4\pi r_{\rm em}^2c=2.7\times10^{-4}~{\rm erg~cm^{-3}}(L_{\rm iso}/10^{28}~{\rm erg~s^{-1}})(r_{\rm em}/10^{10}~{\rm cm})^{-2}$. Here, we consider the highest observed isotropic radio luminosity of ILTJ1101+5521 is $L_{\rm iso}\sim 10^{28}~{\rm erg~s^{-1}}$ \citep{deRuiter25}.

For a viable emission process, two conditions must be met. First, the total energy density of the source plasma must exceed the radiated energy density by a factor of $\eta_{\rm rad}^{-1}$, where $\eta_{\rm rad}$ is the radiation efficiency. Second, for this plasma to be magnetically confined, the magnetic energy density, $U_B$, must dominate the plasma energy density. This imposes a much stronger condition: $U_B \gtrsim \zeta U_{\rm EM}/\eta_{\rm rad}$, where $\zeta \gg 1$ is a factor quantifying the degree of magnetic dominance. This requirement leads to a direct constraint on the emission:
\begin{align} 
r_{\rm em}&\lesssim \fraction{\mu_s^2\eta_{\rm rad}c}{2\zeta L_{\rm iso}}{1/4}\simeq3.5\times10^{12}~{\rm cm}\fraction{\eta_{\rm rad}}{\zeta}{1/4}\nonumber\\ 
&\times\fraction{\mu_s}{10^{34}~{\rm G~cm^3}}{1/2}\fraction{L_{\rm iso}}{10^{28}~{\rm erg~s^{-1}}}{-1/4}.\label{rem}
\end{align} 
The above upper limit is much larger than the orbital separation $a\sim10^{10}~{\rm cm}$, which means that there is enough magnetic energy driving LPTs in the magnetosphere of the binary system.
This is also consistent with the magnetic mirror picture that produces the LCDM, see Section \ref{mechanism}.

The small pulse duty cycles observed in most LPTs, typically in the range of $\Delta t/P \sim (10^{-3}-10^{-1})$ \citep{McSweeney25}, strongly suggest a geometric origin for the pulse profile. This naturally leads to the ``lighthouse effect'', analogous to that of radio pulsars. Within this framework, the beaming angle can be constrained by the observed pulse duty cycle,
\be 
\Theta_{\rm obs}=\frac{2\pi\Delta t}{P}\simeq0.1~{\rm rad}\fractionz{\Delta t}{100~{\rm s}}\fraction{P}{100~{\rm min}}{-1}.
\ee 
If the radiation is directed along the magnetic field lines or at a specific angle to them (as in the LCDM discussed in Section \ref{mechanism}), the resulting beaming angle is determined by the global geometry of the magnetosphere. This principle holds true for both the unipolar-inductor and the magnetosphere-interaction mechanisms, as the emission in both scenarios depend on the structured magnetic field connecting the two stars (see Figures~\ref{UI} and \ref{MI}). Besides, the presence of such a beaming effect has a significant implication for population statistics: the intrinsic number density of the LPT population must be higher than the observed number by a correction factor, see Section \ref{density} in detail. In the following subsections, we will first analyze the specific field beaming effects expected for the unipolar-inductor and the magnetosphere-interaction mechanisms, respectively.

\subsubsection{Field beaming of unipolar-inductor mechanism} 

We first analyze the field beaming of the unipolar-inductor mechanism, the geometry of which is depicted in Figure~\ref{UI}.
We consider that the magnetic field of the WD is dipole on a large scale. In a standard polar coordinate system $(r, \theta)$, the magnetic field lines are described by the relation
\be 
r=R_{\max}\sin^2\theta,
\ee
where $R_{\max}$ denotes the maximum distance at which the field line crosses the magnetic equator. Define the angle between the magnetic axis and the magnetic field at $(r,\theta)$ as $\psi$, one has 
\be 
\psi=\theta+\arccos\left(\frac{2\cos\theta}{\sqrt{1+3\cos^2\theta}}\right),
\ee 
and its corresponding difference reads
\be 
\frac{d\psi}{d\theta}=\frac{3(1+\cos^2\theta)}{1+3\cos^2\theta}.\label{dpsi}
\ee
The polar angle difference at the WD's surface is
\begin{align}
\delta\theta_{s,\max}&=\arcsin\sqrt{\frac{R_s}{a-R_c}}-\arcsin\sqrt{\frac{R_s}{a+R_c}}\nonumber\\ 
&\simeq\frac{R_c}{a}\fraction{R_s}{a}{1/2}~~~{\rm for}~a\gg R_c.
\end{align}
Thus, if the emission region is near the surface of the WD with $\theta\ll1$, using Eq.(\ref{dpsi}), the field beaming angle is
\begin{align}
\Theta_b&<\delta\psi_{\max}=\frac{3R_c}{2a}\fraction{R_s}{a}{1/2}\simeq0.04~{\rm rad}\fraction{P}{100~{\rm min}}{-1}\nonumber\\ 
&\times\fraction{M_s+M_c}{M_\odot}{-1/2}\fractionz{R_c}{10^{10}~{\rm cm}}\fraction{R_s}{10^{9}~{\rm cm}}{1/2}. 
\end{align} 
It is crucial to recognize that $\delta\psi_{\max}$ represents a strict upper limit on the field beaming angle, $\Theta_b$. This is because the unipolar-inductor circuit inherently involves counter-streaming currents, i.e., an upward current along one set of field lines and a downward current along another (as depicted in Figure~\ref{UI}). 
If the radiation mechanism is tied to the current direction, this geometry would produce two oppositely directed emission beams for any given flux tube, effectively halving the observable beam width compared to the total angular extent of the active region. 
This theoretically derived upper limit on the beaming angle, $\Theta_b \lesssim \delta\psi_{\max}$. Some LPTs exhibit duty cycles that imply a much larger beaming angle, on the order of $\Theta_{\rm obs} \sim 0.1$ rad. Thus, an emission region near the surface of the WD cannot naturally produce such a relatively wide beam. This discrepancy strongly suggests that the emission from LPTs must originate from a much larger, higher-altitude region within the magnetosphere, rather than from the low-altitude polar caps. 

The calculation of the field beaming angle becomes more complex for emission regions located at high altitudes, far from the WD's surface. In this regime, the field beaming angle is determined by the angular extent of the plasma-filled magnetic flux tube at the emission altitude. We first give the approximate field line length, $l$, of a given dipole magnetic field line,
\be
l\simeq \frac{1}{2}R_{\max}(1-\cos\theta)(3+\cos\theta),
\ee
for $\theta\lesssim1$. 
where the empirical approximation of $\sqrt{1+3\cos^2\theta}\sim1+\cos\theta$ is used to calculate the field line length. 
Assuming that the velocities of the plasma flows at different field lines are the same, the outer boundary of the magnetic flux tube should have the same length from the emission region to the WD center.
We consider two lines with $R_{\max,1}=a-R_c$ and $R_{\max,2}=a+R_c$, respectively, see Figure \ref{UI}. For these two lines with the same length $l_1(\theta_1)=l_2(\theta_2)$ and $\theta_2=\theta$ and $\theta_1=\theta+\delta\theta$, we obtain
\be 
\frac{(1-\cos\theta)(3+\cos\theta)}{[1-\cos(\theta+\delta\theta)][3+\cos(\theta+\delta\theta)]}=\frac{a-R_c}{a+R_c}.
\ee 
The polar angle difference $\delta\theta$ at the emission region is solved as 
\be 
\delta\theta\simeq\frac{R_c}{a}\frac{(1-\cos\theta)(3+\cos\theta)}{\sin\theta(1+\cos\theta)}\sim\frac{R_c}{a},\label{dtheta}
\ee 
Combing Eq.(\ref{dpsi}) and Eq.(\ref{dtheta}), the field beaming angle is given by
\begin{align}
\Theta_b &<\delta\psi_{\max}=\frac{3(1+\cos^2\theta)}{1+3\cos^2\theta}\delta\theta
=f_\theta\frac{R_c}{a}\simeq0.2~{\rm rad}f_\theta\nonumber\\
&\times\fraction{P}{100~{\rm min}}{-2/3}\fraction{M_s+M_c}{M_\odot}{-1/3}\fractionz{R_c}{10^{10}~{\rm cm}}\label{thetab1}
\end{align}
with a $\theta$-dependent factor of 
\be 
f_\theta\equiv\frac{3\sin\theta(1+\cos^2\theta)(3+\cos\theta)}{(1+3\cos^2\theta)(1+\cos\theta)^2}.\label{ftheta} 
\ee 
Such a constraint is consistent with the observed duty cycle of LPTs, $\Delta t/P\sim(10^{-3}-10^{-1})$. Thus, the emission region of LPTs seems to placed at relatively high latitudes for the unipolar-inductor mechanism.

\subsubsection{Field beaming of magnetic reconnection}

Next, we analyze the field beaming for the magnetosphere-interaction mechanism. We model the reconnection region as a current sheet with a characteristic length $L$, width $W$, and thickness $H$. In this scenario, particles are accelerated out of a classic ``X-point'' magnetic geometry, as depicted in Figure~\ref{MI}. Plasma flows into this region with a velocity $v_{\rm in}$ and density $\rho_{\rm in}$ over an area $L W$, resulting in a mass inflow rate of $\Phi_{\rm in} \simeq \rho_{\rm in} v_{\rm in} LW$. The accelerated plasma is then ejected as an outflow with velocity $v_{\rm out}$ and density $\rho_{\rm out}$ through the exhaust channel of area $H W$, giving a mass outflow rate of $\Phi_{\rm out} \simeq \rho_{\rm out} v_{\rm out} HW$. By invoking mass conservation, $\Phi_{\rm in} \simeq \Phi_{\rm out}$, we can solve for the thickness of the reconnection region:
\be 
H\sim \frac{v_{\rm in}}{v_{\rm out}}L= M_AL\sim(0.01-0.1)L,
\ee 
for a rough assumption of incompressible flow with $\rho_{\rm in}\sim\rho_{\rm out}$.  
Accelerated electrons in the outflow move from the reconnection region to the polar region of the WD and further produce LPTs, see Section \ref{mechanism} and Figure \ref{losscone} for the detailed discussion. 
Similar to the case of the unipolar-inductor mechanism, we consider two lines with $R_{\max,1}=R_{\rm rec}$ and $R_{\max,2}=R_{\rm rec}+H$, respectively.
According to Eq.(\ref{Rrec}), Eq.(\ref{thetab1}) and Eq.(\ref{ftheta}), the field beaming angle for the magnetosphere interaction is 
\begin{align} 
\Theta_b=f_\theta\frac{H}{R_{\rm rec}}\simeq(0.01-0.1)~{\rm rad}f_\theta\fractionz{L}{a}\left[\fraction{\mu_c}{\mu_s}{1/3}+1\right]. 
\end{align}
Such a beaming angle from the magnetic reconnection is also consistent with the observed duty cycle of the LPTs. 

\subsection{Radiation mechanism}\label{mechanism}

\begin{figure}
    \centering
    \includegraphics[width = 0.85\linewidth, trim = 30 150 0 0, clip]{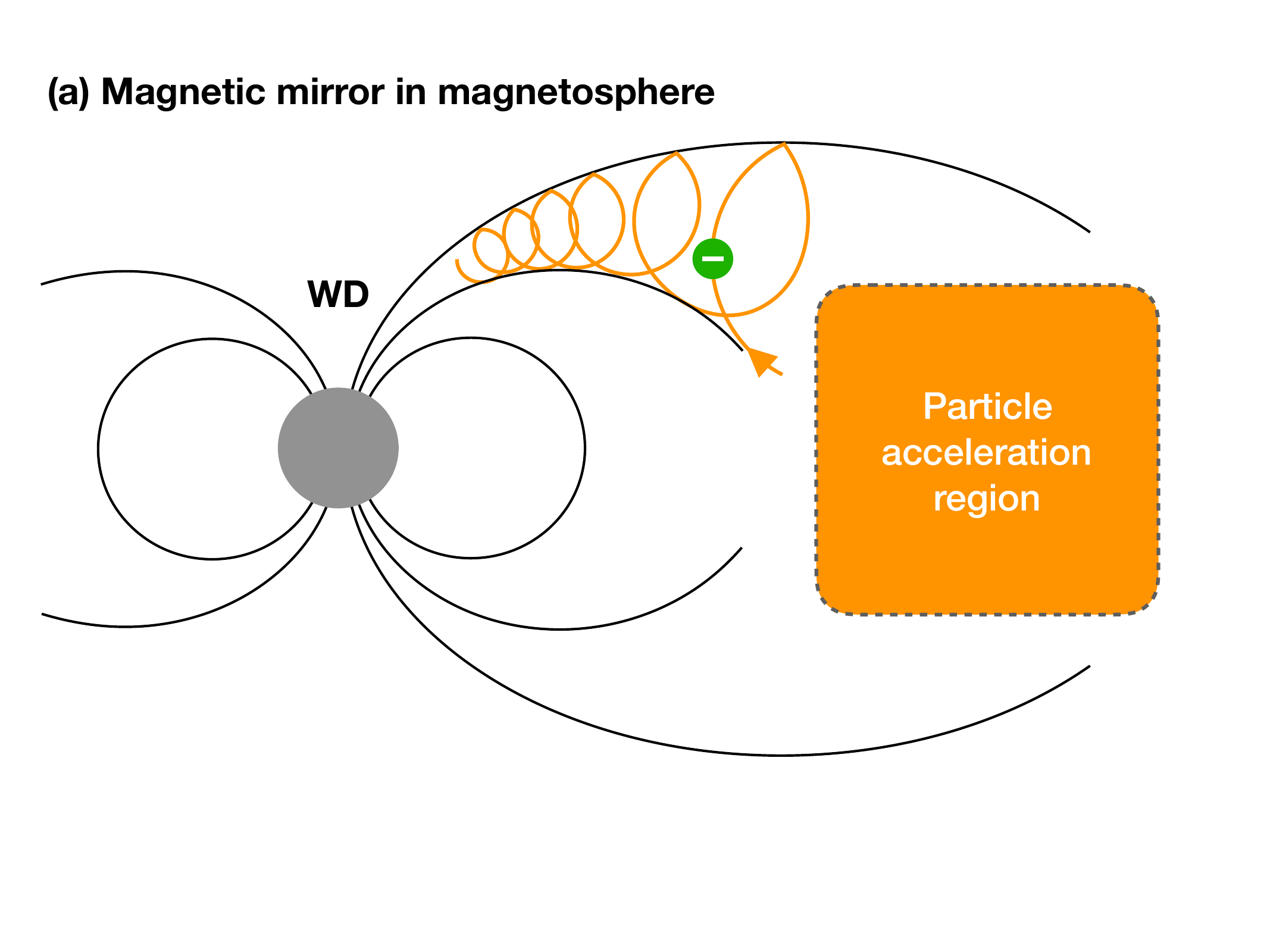} 
    \includegraphics[width = 0.85\linewidth, trim = 30 200 0 0, clip]{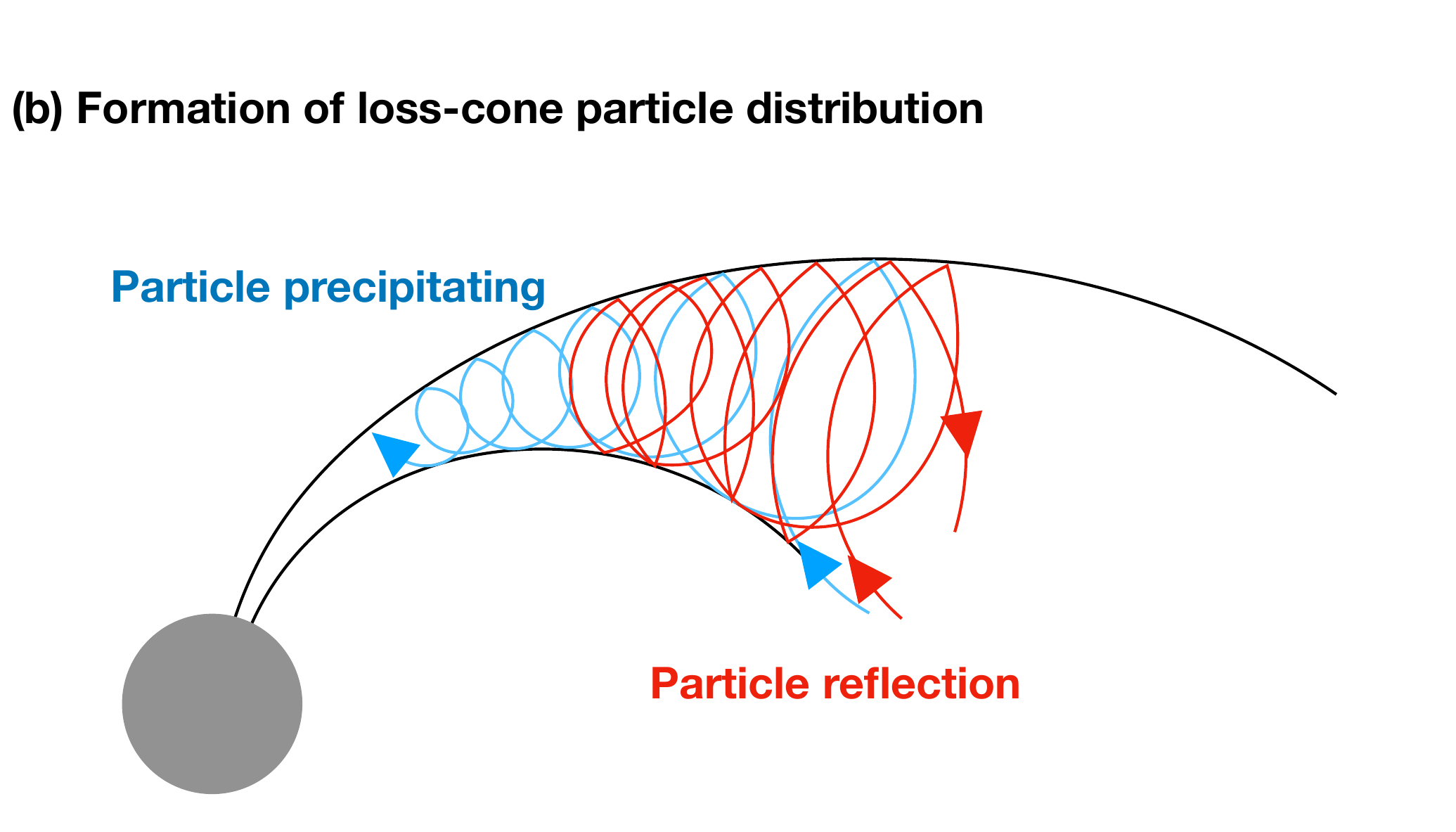} 
    \includegraphics[width = 0.85\linewidth, trim = 30 150 0 0, clip]{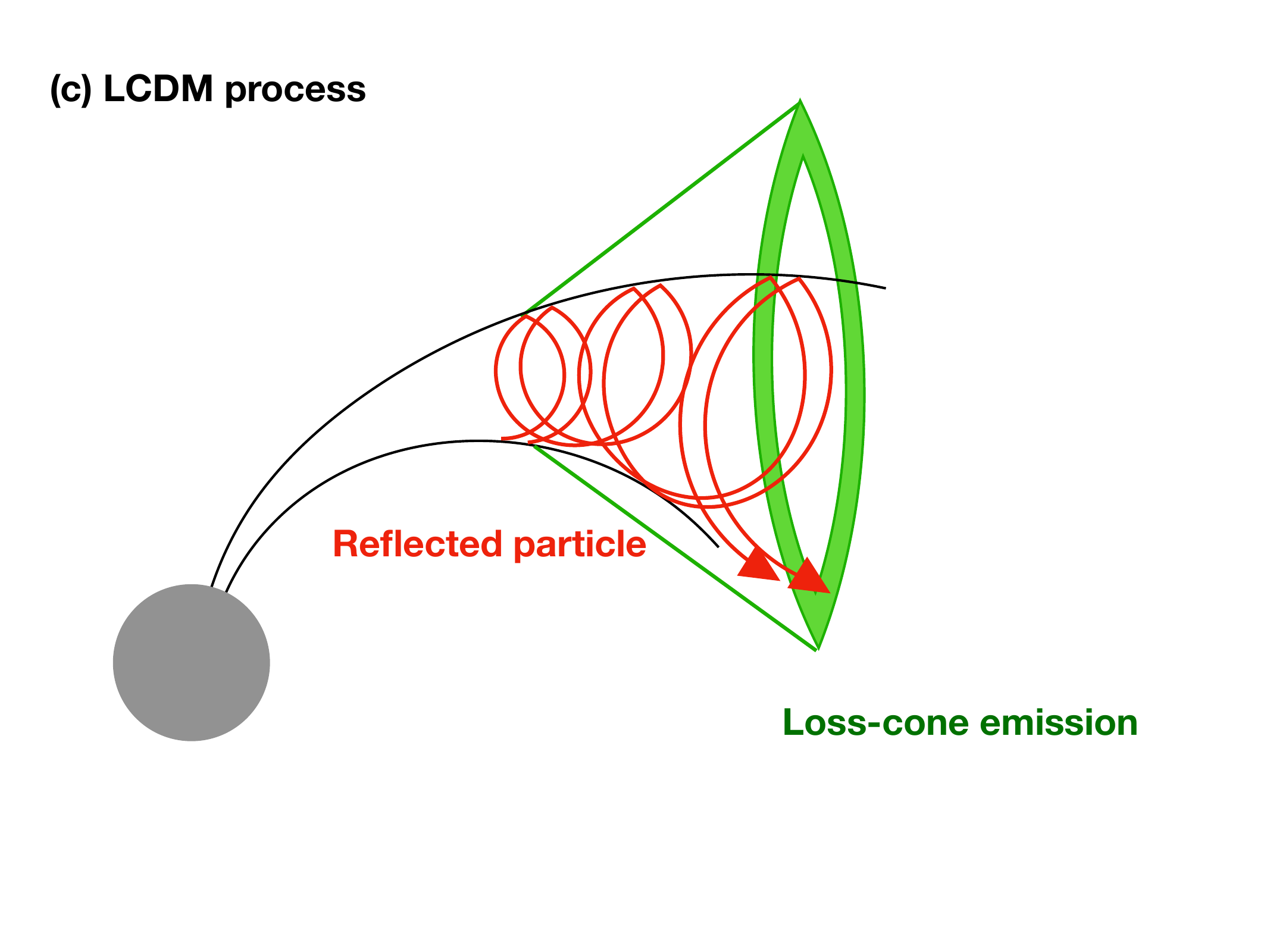} 
    \caption{Schematic of the loss-cone-driven maser (LCDM) process within the magnetosphere of a WD--MD binary. The process begins with electrons being accelerated, either at the MD's surface (in the unipolar-inductor mechanism) or in a magnetic reconnection region (in the magnetosphere-interaction mechanism). (a) These energetic electrons are injected onto magnetic field lines and propagate towards the WD. (b) As they encounter the strengthening magnetic field ($\nabla\bm{B} \parallel \bm{B}$), the magnetic mirror effect reflects electrons with large pitch angles, while allowing those with small pitch angles to precipitate. (c) This loss-cone anisotropy provides the free energy to power LCDM, generating coherent radio waves.
    }\label{losscone} 
\end{figure}

In this section, we focus on the loss-cone-driven maser \citep[LCDM][]{Wu79,Hewitt82,Melrose82,Melrose84} as the primary radiation mechanism of LPTs in WD--MD binaries. This radiation mechanism, a specific form of electron cyclotron maser emission (ECME) extensively developed by \citet{Melrose82}, is particularly well-suited for our scenario because the large-scale magnetic field geometry of a WD--MD binary naturally forms the ``magnetic mirrors'' required for its operation (see Figure \ref{losscone}). Within the LCDM framework, both the high brightness temperatures and the intrinsic beaming of LPTs can be naturally explained as follows.
The interaction between electromagnetic waves is characterized by a frequency $\omega$, wavevector $\bm{k}$, and a dispersion relation, and a population of electrons described by a momentum distribution function $f(\bm{p})$, where $\bm{p} = \gamma m_e \bm{v}$. A fundamental requirement for ECME to occur is the satisfaction of the gyromagnetic resonance condition:
\be 
\omega-\frac{s\omega_B}{\gamma}-k_\parallel v_\parallel=0,\label{resonance}
\ee
where $\omega_B=eB/m_ec$ is the electron-cyclotron frequency, $\parallel$ and $\perp$ denote the components relative to the $\bm{B}$, $s$ denotes the harmonic number of resonance. 

The LCDM mechanism requires a specific velocity-space anisotropy known as a ``loss cone'', which describes a particle distribution that is depleted of particles at small pitch angles ($\alpha = \arctan(v_\perp/v_\parallel)$).
This depletion of particles at small pitch angles in the magnetic mirror creates a positive gradient in the velocity distribution perpendicular to the magnetic field. Specifically, as $v_\perp$ increases from zero, the number of particles, represented by the distribution function $f$, initially rises. This results in a region where $\partial f/\partial v_\perp > 0$, a condition known as a population inversion, which is the fundamental requirement for maser amplification.
This loss-cone distribution can be naturally formed within the WD--MD binary magnetosphere. 
We consider the plasma in the magnetosphere between the WD and the MD to be dominated by an electron-ion medium. Given that electrons are much lighter, the coherent radiation is expected to be primarily governed by the electrons.
The process begins with the acceleration of energetic electrons, either near the MD's surface via the unipolar-inductor mechanism or within the magnetosphere-interaction reconnection region. These newly accelerated electrons are then injected onto the magnetic field lines connecting the two stars and propagate towards the WD (panel (a) of Figure \ref{losscone}). As they travel along these converging field lines into the strengthening magnetic field ($\nabla\bm{B} \parallel \bm{B}$) near the WD, the magnetic mirror effect comes into play. Electrons with large pitch angles are reflected back, while those with small pitch angles are able to penetrate the mirror and precipitate onto the WD surface, thereby creating the required loss-cone anisotropy (panel (b) of Figure \ref{losscone}).
We consider that the magnetic field strength of a magnetic mirror at the midplane is $B_0$ and the maximum field strength of the magnetic mirror is $B_{\max}$, then the smallest pitch angle $\alpha_m$ of a confined electron is given by
\be 
\sin^2\alpha_m=\frac{B_0}{B_{\max}}.
\ee 
This defines the boundary of a region in momentum space $f(\bm{p})$ in the shape of a loss cone. Because a part of electrons with small pitch angles precipitate, a plasma held by a magnetic mirror is always anisotropic and never isotropic. In the picture of the magnetic WD -- MD binary, $B_{\max}$ corresponds to the field strength near the WD, $B_{\max}\lesssim B_s$, and $B_0$ corresponds to the field strength of the magnetosphere at the large-scale region, $B_0\sim\mu_s/r^3$. Thus, the pitch angle of electrons is given by
\begin{align} 
\alpha_m&\gtrsim \fraction{R_s}{r}{3/2}\nonumber\\ 
&\simeq0.03~{\rm rad}\fraction{R_s}{10^9~{\rm cm}}{3/2}\fraction{r}{10^{10}~{\rm cm}}{-3/2}.
\end{align} 
Therefore, the magnetic mirror effect creates a one-sided, unstable distribution of reflected electrons, which in turn drives the maser. 

Because $\omega_p / \omega_B \ll 1$ under typical parameters, where $\omega_p=(4\pi e^2n_e/m_e)^{1/2}$ is the plasma frequency of electrons and $\omega_B=eB/m_ec$ is the electron cyclotron frequency, the LCDM in this scenario is not expected to be significantly suppressed \citep[see the details in][]{Hewitt82,Melrose84}.
Therefore, the growth rate of the LCDM is approximately given by \citep{Melrose82}
\be 
\frac{\Gamma_s}{\omega}\simeq\eta_s\frac{n_0}{n_e}\fraction{\omega_p}{\omega}{2}\fraction{c}{v_0}{2}\left(\frac{v_0}{c}\sin\alpha_0\cos\alpha_0\right)^{2s-2},\label{Gammas}
\ee 
where $v_0\equiv k_\parallel c^2/\omega$, $n_0$ is the number density of electrons with $v>v_0$, $\eta_s$ is a number typically of order unity, and $\alpha_0$ is the tangent angle with resonant circle that is defined by Eq.(\ref{resonance}), see Figure 3 of \citet{Melrose82}. In particular, we apply the growth rate of the WD -- MD scenario, and one has 
\begin{align}
\Gamma&=\Gamma_1\simeq\frac{4\pi e^2n_0}{m_e\omega}\fraction{c}{v_0}{2}\simeq5.0\times10^4~{\rm s^{-1}}\nonumber\\ 
&\times\fractionz{n_0}{10^2~{\rm cm^{-3}}}\fraction{v_0}{0.1c}{-2}\fraction{\nu}{100~{\rm MHz}}{-1},
\end{align} 
for $s=1$ and $\eta_s\sim1$. The value of $n_0$ is poorly known, which depends on the specific form of distribution of the electron energy/velocity; meanwhile, it should be less than the total electron number density $n_e$ estimated by Eq.(\ref{edensity}).

Assuming that the particle distribution function satisfies $f\propto v^{-a}$, for $s<(a+1)/2$, the maximum growth rate occurs when $v_0=v_{\min}/(1-\sin\alpha_0)$ \citep{Melrose82}. 
We define $\theta$ as the wave angle, i.e., $\cos\theta=k_\parallel/k$.
Using the approximation of $k\simeq\omega/c$, the maximum growth rate occurs at a wave angle of $\theta_0$ with resonant
\be 
\cos\theta_0\simeq \frac{v_0}{c},\label{theta0}
\ee
and at a frequency of 
\be 
\omega=\left[1+\frac{1}{2}\fraction{v_0\cos\alpha_0}{c}{2}\right]s\omega_B.\label{ecmefreq}
\ee 
For $s=1$ and $v_0\lesssim c$, the emission frequency is about the electron cyclotron frequency at the emission region 
\begin{align} 
\nu&\simeq\nu_B\sim\frac{e\mu_s}{2\pi m_ec r^3}\simeq230~{\rm MHz}\fractionz{\mu_s}{10^{34}~{\rm G~cm^3}}\nonumber\\ 
&\times\fraction{P}{100~{\rm min}}{-2}\fraction{M_s+M_c}{M_\odot}{-1}\fraction{r}{a}{-3}.
\end{align} 
This is consistent with the observed typical frequency of LPTs.
Based on Eq.(\ref{ecmefreq}), specially for $v_0\lesssim c$, the relative frequency bandwidth near the maximum growth rate is about
\be 
\frac{\delta \omega}{\omega}\sim\fraction{v_0}{c}{2}\alpha_0\delta\alpha\lesssim1,\label{domega}
\ee 
where $\delta\alpha$ is defined by the particle distribution $f$ and $\alpha_0$ (with $f=0$ for $\alpha<\alpha_0-\delta\alpha$ and $f$ rises linearly with $\alpha$ over the range of $\alpha_0-\delta\alpha<\alpha<\alpha_0$).
This result suggests that the radiation from the LCDM has a relatively narrow spectrum. For a certain frequency $\omega$, the angular range near the maximum growth rate is about 
\be 
\delta\cos\theta\simeq\frac{\delta v_0}{c}\sim\frac{v_0}{c}\delta\alpha\lesssim1.\label{dtheta0}
\ee
Eq.(\ref{theta0}) and Eq.(\ref{dtheta0}) suggest that the LCDM has a cone-beaming structure see panel (c) of Figure \ref{losscone}.

Furthermore, the full saturation of LCDM occurs at an energy density of 
\be 
U_{\rm free}\simeq n_0m_ev_0^2\alpha_0^3.
\ee 
This is equivalent to the electron energy density that would exist if the loss cone was fully filled. 
The maser reaches saturation once a significant portion of its available free energy has been transformed into radiation. The brightness temperature can be estimated by $k_BT_B(\omega/2\pi c)^3(\delta\omega/\omega)\delta\Omega\sim U_{\rm free}$, where $\delta\Omega=2\pi\delta(\cos\theta)$ is the solid angle filled by the radiation, leading to an allowed maximum brightness temperature of of substructure\footnote{In Eq.(16) of \citet{Melrose82}, a power index of $3$ of the factor of $(2\pi c^2/\omega v_0)$ was omitted.}
\be 
T_{B,{\max}}\simeq\frac{n_0m_ev_0^2}{2\pi k_B}\fraction{2\pi c^2}{\omega v_0}{3}\fraction{\alpha_0}{\delta\alpha}{2},
\ee 
where the approximations of Eq.(\ref{domega}) and Eq.(\ref{dtheta0}) are used in the above result. 

For electrons in a magnetic mirror with a typical length of $L$, the saturation of the maser for operating steadily will reduce the above maximum enhancement by the factor $v_0/\Gamma L$ due to the finite rate at which the loss cone can empty. 
The maximum power of the maser is constrained because if the pitch angle diffusion timescale $\alpha_0^2/D_{\alpha\alpha}\sim 1/\Gamma$ ($D_{\alpha\alpha}\sim(\delta\alpha)^2\Gamma$ is the pitch angle diffusion coefficient) becomes faster than the particle bounce timescale $L/v_0$ within the magnetic trap, the loss cone will become filled. This filling eliminates the necessary anisotropy, thereby preventing any further wave growth and limiting the maser's output.
In conclusion, a dynamic equilibrium is established where the continuous reflection at the magnetic mirror sustains this unstable distribution, while the maser emission itself scatters particles and works to smooth it out. Consequently, the maser operates in a self-regulating manner near its emission threshold.
In this case, although the electrons' acceleration continues, the maser should operate sporadically near the threshold level, suppressing the observed effective flux. 
Meanwhile, such a picture further predicts that the light curve of LPTs should have some substructure with a duty cycle of $f_{\rm cyc}=v_0/\Gamma L$.
Thus, the suppressed brightness temperature may be estimated by
\begin{align} 
T_B&\simeq\frac{\pi m_e^2c^4v_0^2}{e^2k_BL\omega^2}\simeq3.1\times10^{13}~{\rm K}\fraction{v_0}{0.1c}{2}\fraction{\nu}{100~{\rm MHz}}{-2}\nonumber\\ 
&\times\fraction{P}{100~{\rm min}}{-2/3}\fraction{M_s+M_c}{M_\odot}{-1/3}\fraction{L}{a}{-1},
\end{align}
for $s=1$, $\eta_s\sim1$ and $\delta\alpha\sim\alpha_0$,
Notice that the corrected bright temperature is independent of $n_0$ which is defined by the electron number density with the condition of $v>v_0$.
This is consistent with the observed brightness temperature of LPTs constrained by Eq.(\ref{brightness}).

Finally, we briefly discuss the effect of ions on the LCDM. Because the mass of ions is much larger than that of electrons, their cyclotron frequency is much lower than the electron cyclotron frequency, which prevents them from resonantly interacting with the high-frequency ECME waves. As analyzed in the previous works, ions are treated as a passive background providing charge neutrality, without directly participating in the (ECME) process \citep[e.g.,][]{Wu79,Hewitt82,Melrose82,Melrose84}. Recently, further studies revealed that ions play an indirect role by being accelerated out of the ECME source region by electric fields to form the low-density plasma cavity essential for the maser, while their collective instabilities generate larger ion holes that act as potential barriers to trap and structure the actual electron-hole emission sources \citep[see the details in the review of][]{Treumann06}. The non-linear dynamics of this coupled ion-electron system are complex, and a detailed treatment is beyond the scope of this paper.   

\section{Propagation effect: Faraday conversion}\label{conversion}

A striking observational feature of LPTs is their complex polarization behaviors, which include both high degrees of circular polarization and instances of polarization mode conversion \citep[e.g.,][]{Caleb24, Men25}. Explaining these properties requires considering the propagation effects within the binary's magnetosphere. 
Specifically, in such a strongly magnetized environment, the local electron cyclotron frequency can become comparable to the observing frequency. 
Faraday conversion becomes an important propagation effect, profoundly altering the polarization state of the escaping radiation. 

We first summarize the theory of polarized radiative transfer in a cold, magnetized electron-ion plasma \citep[e.g.,][]{Melrose10,Gruzinov19,Qu23,Xia23}.
In this case, the refractive index is given by the Altar-Appleton-Hartree dispersion relation \citep[e.g.,][]{Stix92}, 
\be 
n^2 \equiv\frac{k^2c^2}{\omega^2}= 1 - \frac{2\omega_{p}^2(\omega^2 - \omega_{p}^2)/\omega^2}{2(\omega^2 - \omega_{p}^2) - \omega_B^2 \sin^2 \theta \pm \omega_B \Delta},\label{AAH} 
\ee
where
\be 
\Delta = \left[ \omega_B^2 \sin^4 \theta + 4\omega^{-2}(\omega^2 - \omega_{p}^2)^2 \cos^2 \theta \right]^{1/2},\label{Delta} 
\ee
and $\theta$ is the angle between the line of sight and the field direction. 
The two natural modes depend on the $\pm$ choice of sign in Eq.(\ref{AAH}), and the conditions of Quasi-Transverse (QT) and Quasi-Longitudinal (QL) regimes depend on the relative size of two terms in $\Delta$ of Eq.(\ref{Delta}), i.e., $\omega_B^2 \sin^4 \theta \gtrsim 4\omega^{-2}(\omega^2 - \omega_{p}^2)^2 \cos^2 \theta$ for QT and $\omega_B^2 \sin^4 \theta \lesssim 4\omega^{-2}(\omega^2 - \omega_{p}^2)^2 \cos^2 \theta$ for QL. 

For the magnetosphere environment of a WD--MD binary with typical parameters, we are mainly interested in the following two cases: 1) weak-field regime with $\omega_p\lesssim\omega_B\lesssim\omega$; 2) strong-field regime with $\omega_p\lesssim\omega\lesssim\omega_B$. 
The conditions of QT and QL are approximately $\omega_B \sin^2 \theta \gtrsim 2\omega |\cos\theta|$ and $\omega_B \sin^2 \theta \lesssim 2\omega |\cos\theta|$, respectively.
The difference of the wavevectors of two natural modes is $\Delta k=|n_+-n_-|\omega/c$, where $n_+$ and $n_-$ are refractive indexes of two natural modes. Thus, in the weak-field regime, one approximately has
\be
\Delta k\simeq\frac{\omega_p^2\omega_B}{c\omega^2}\left\{
\begin{aligned} 
&|\cos\theta|, &  {\rm QL}\\
&\frac{\omega_B}{2\omega}\sin^2\theta, & {\rm QT}
\end{aligned}\right.\label{dk}
\ee
And in the strong-field regime, one approximately has
\be
\Delta k\simeq\frac{\omega_p^2}{c\omega}\left\{
\begin{aligned} 
&\frac{\omega}{\omega_B}|\cos\theta|^{-1}, &  {\rm QL}\\
&\frac{1}{2}. & {\rm QT}
\end{aligned}\right.\label{dk1}
\ee

The evolution of a wave's polarization is governed by the two natural modes of the magnetized plasma, which are determined by the local plasma conditions, specifically the electron number density, $n_e$, and the magnetic field vector, $\bm{B}$. This evolution can be described by the following radiative transfer equation \citep{Melrose10}:
\begin{equation}
    \frac{d \boldsymbol{P}}{d z}=\boldsymbol{\rho} \times \boldsymbol{P},  \label{dPdz}
\end{equation}
Here, we simply assume that the wave is fully polarized and propagates along the $z$-direction. 
The polarization vector is defined by $\bm{P}=(Q,U,V)$ that describes the polarization state and the direction of the two natural modes in the Poincar\'{e} sphere is
\be 
\boldsymbol{\rho}\equiv \left(\begin{array}{l} 
        \rho_Q \\
        \rho_U \\
        \rho_V
    \end{array}\right)=\left(\begin{array}{l}
        -\Delta k\cos 2\chi_B\cos2\psi_B \\
        -\Delta k\cos 2\chi_B\sin2\psi_B \\
        -\Delta k\sin 2\chi_B
        \end{array}\right),
\ee 
where $2\chi_B$ and $2\psi_B$ denote the latitude and longitude of the natural mode on the Poincar\'{e} sphere, respectively, which also correspond to the ellipticity and the orientation of the natural mode, respectively. 

Since the radiation mechanism of LPTs is proposed to operate via the LCDM process, with an emission frequency $\omega \sim \omega_B$ in Section \ref{mechanism}, both weak-field and strong-field regimes are possible. The former corresponds to the case where the radiation direction deviates from the WD/MD, while the latter corresponds to the case where it is close to the WD/MD. In the following, we discuss the observable polarization characteristics in the weak-field and strong-field regimes, respectively:

1) The weak-field regime with $\omega_p\ll\omega_B\ll\omega$:
In the QL region with $2\chi_B\simeq\pi/2$, one has 
\be 
|\rho_V|\simeq \frac{\omega_p^2\omega_B}{c\omega^2}|\cos\theta|.
\ee 
In this case, the classical Faraday rotation effect is dominant and it becomes significant when the angle of $\theta_{\rm FR}\equiv \left|\int \rho_Vdz\right|\simeq\rho_VL$ becomes larger than a typical value of $\sim 2\pi$. Thus, we can rewrite $\theta_{\rm FR}$ as
\be 
\theta_{\rm FR}=2\pi\fraction{\nu_{\rm FR}}{\nu}{2}
\ee 
with
\be 
\nu_{\rm FR}=\frac{\omega_{\rm FR}}{2\pi}\equiv\frac{1}{2\pi}\fraction{\omega_p^2\omega_BL\cos\theta}{2\pi c}{1/2}.
\ee 
On the other hand, in the QT region with $2\chi_B\simeq0$, one has 
\be 
|\rho_L|\equiv|\rho_Q+i\rho_U |\simeq \frac{\omega_p^2\omega_B^2}{2c\omega^3}\sin^2\theta.
\ee 
Similarly, the Faraday conversion effect is significant when the angle of $\theta_{\rm FC}\equiv \left|\int \rho_Ldz\right|$ becomes larger than $\sim 2\pi$. 
Similar, we can rewrite $\theta_{\rm FC}$ as
\be 
\theta_{\rm FC}=2\pi\fraction{\nu_{\rm FC}}{\nu}{3},
\ee
with 
\be 
\nu_{\rm FC}=\frac{\omega_{\rm FC}}{2\pi}\equiv\frac{1}{2\pi}\fraction{\omega_p^2\omega_B^2L\sin^2\theta}{4\pi c}{1/3}.\label{nufw}
\ee 

2) The strong-field regime with $\omega_p\ll\omega\ll\omega_B$: In the QL region with $2\chi_B\simeq\pi/2$, the wavevector differenve is independent of the wave frequency. In this case, there is no observable effects. 
In the QT region with $2\chi_B\simeq0$, one has
\be 
|\rho_L|=\frac{\omega_p^2}{2c\omega}.
\ee 
Similarly, the conversion angle is
\be 
\theta_{\rm FC}=2\pi\fractionz{\nu_{\rm FC}}{\nu},
\ee
with 
\be 
\nu_{\rm FC}=\frac{\omega_{\rm FC}}{2\pi}\equiv\frac{\omega_p^2L}{8\pi^2 c}. \label{nufs}
\ee 

In the magnetosphere of the WD -- MD binary, both the condition of $\nu_B\gtrsim \nu_{\rm obs}$ and $\nu_B\lesssim \nu_{\rm obs}$ are possible, corresponding to the line of sight deviating from or close to the WD/MD, respectively. 
For the weak-field (Eq.(\ref{nufw})) and strong-field (Eq.(\ref{nufs})) regimes, the corresponding critical frequencies are estimated by
\begin{align} 
&\nu_{\rm FC}\simeq330~{\rm MHz}~\fraction{\sin\theta}{10^{-2}}{2/3}\fraction{n_e}{10^5~{\rm cm^{-3}}}{1/3}\fraction{L}{a}{-5/3}\nonumber\\ 
&\times\fraction{\mu_s}{10^{34}~{\rm G~cm^3}}{2/3}\fraction{P}{100~{\rm min}}{-10/9}\fraction{M_s+M_c}{M_\odot}{-5/9},\label{nufc}
\end{align} 
and 
\begin{align} 
\nu_{\rm FC}&\simeq6.6\times10^{12}~{\rm Hz}\fractionz{n_e}{10^5~{\rm cm^{-3}}}\fraction{P}{100~{\rm min}}{2/3}\nonumber\\
&\times\fraction{M_s+M_c}{M_\odot}{1/3}\fractionz{L}{a},\label{nufc1}
\end{align}
respectively,
where $n_e\lesssim10^{5}~{\rm cm^{-3}}$ corresponds to a low accretion rate of $\dot M\lesssim10^{-14}M_\odot{\rm yr^{-1}}$ in the pre-mCV phase, see Eq.(\ref{edensity}), and a small angle of $\theta$ with $\sin\theta\sim10^{-2}$ represents the beaming effect along the line of sight.
The Faraday conversion frequency is comparable to or greater than the observing frequency ($\nu_{\rm FC} \gtrsim \nu$) for the typical parameters. This places the wave propagation squarely in the strong conversion regime. As the radio wave traverses the binary's magnetosphere, its polarization vector will therefore undergo rapid, multiple rotations around the local natural mode axis on the Poincar\'{e} sphere, corresponding to a total conversion angle of $\theta_{\rm FC} \gtrsim 2\pi$. This rapid evolution fundamentally alters the wave's initial polarization, leading to significant changes in both its polarization state (e.g., linear to circular) and its degree of polarization.

While the absolute polarization change of a wave as it traverses the magnetosphere cannot be directly measured, the process leaves a distinct, observable signature in the frequency domain. The Faraday conversion angle, $\theta_{\rm FC}$, is intrinsically frequency-dependent ($\theta_{\rm FC} \propto \nu^{-3}$ for the weak-field regime and $\theta_{\rm FC} \propto \nu^{-1}$ for the strong-field regime). Consequently, the final polarization state and degree measured by an observer will also exhibit a strong dependence on frequency, as the polarization vector $\bm{P}$ will have rotated by different amounts at different frequencies.
This frequency-dependent polarization provides a direct observational test for Faraday conversion. Specifically, we can predict the amplitude of the variation in the linear and circular polarization degrees across a finite observing bandwidth, $\Delta\nu$. 
For a given bandwidth, this variation is approximately given by
\be
\Delta\theta_{\rm FC}\simeq\left\{
\begin{aligned} 
&6\pi\fraction{\nu_{\rm FC}}{\nu}{3}\fractionz{\Delta\nu}{\nu}, ~  &\text{weak field}\\
&2\pi\fractionz{\nu_{\rm FC}}{\nu}\fractionz{\Delta\nu}{\nu}, ~ &\text{strong field}
\end{aligned}\right.
\ee
Thus, in the weak-field regime, the according to Eq.(\ref{nufc}), within a plausible range of magnetospheric parameters, the condition $\nu_{\rm FC} / \nu \sim \text{a few}$ can be satisfied. For a typical radio telescope with a relative bandwidth of $\Delta\nu/\nu \sim 0.5-1$, this translates into a moderate differential Faraday conversion angle of $\Delta\theta_{\rm FC} \sim \text{a few} \times 2\pi$ across the observation band of $\Delta\nu$. Such a large differential rotation is a hallmark of strong Faraday conversion and implies that a distinct, frequency-dependent polarization signature could be observable. Specifically, it predicts a significant conversion between linear and circular polarization states across the observing band\footnote{It should be noted that a more rigorous analysis would require consideration of the initial polarization state, as this also influences the final observed polarization. The calculation presented here is intended as a scaling analysis, designed to estimate the characteristic magnitude of the polarization change rather than to model the precise outcome.}, which is consistent with the observed polarization conversion in the LPT GPM J1839-10 \citep{Men25} and in FRB 20201124A \citep{Xu22}.
On the other hand, in the strong-field regime, the condition $\nu_{\rm FC} \gg \nu$ leads to an extremely large differential Faraday conversion angle, $\Delta\theta_{\rm FC} \gg 2\pi$, across the observation band $\Delta\nu$, resulting in substantial depolarization. This implies that the observed polarization degree of LPTs might be strongly suppressed when the radiation direction approaches alignment with the WD or MD, especially for systems with strong magnetic moments.

\begin{figure}
    \centering
    \includegraphics[width = 1.0\linewidth, trim = 0 0 0 0, clip]{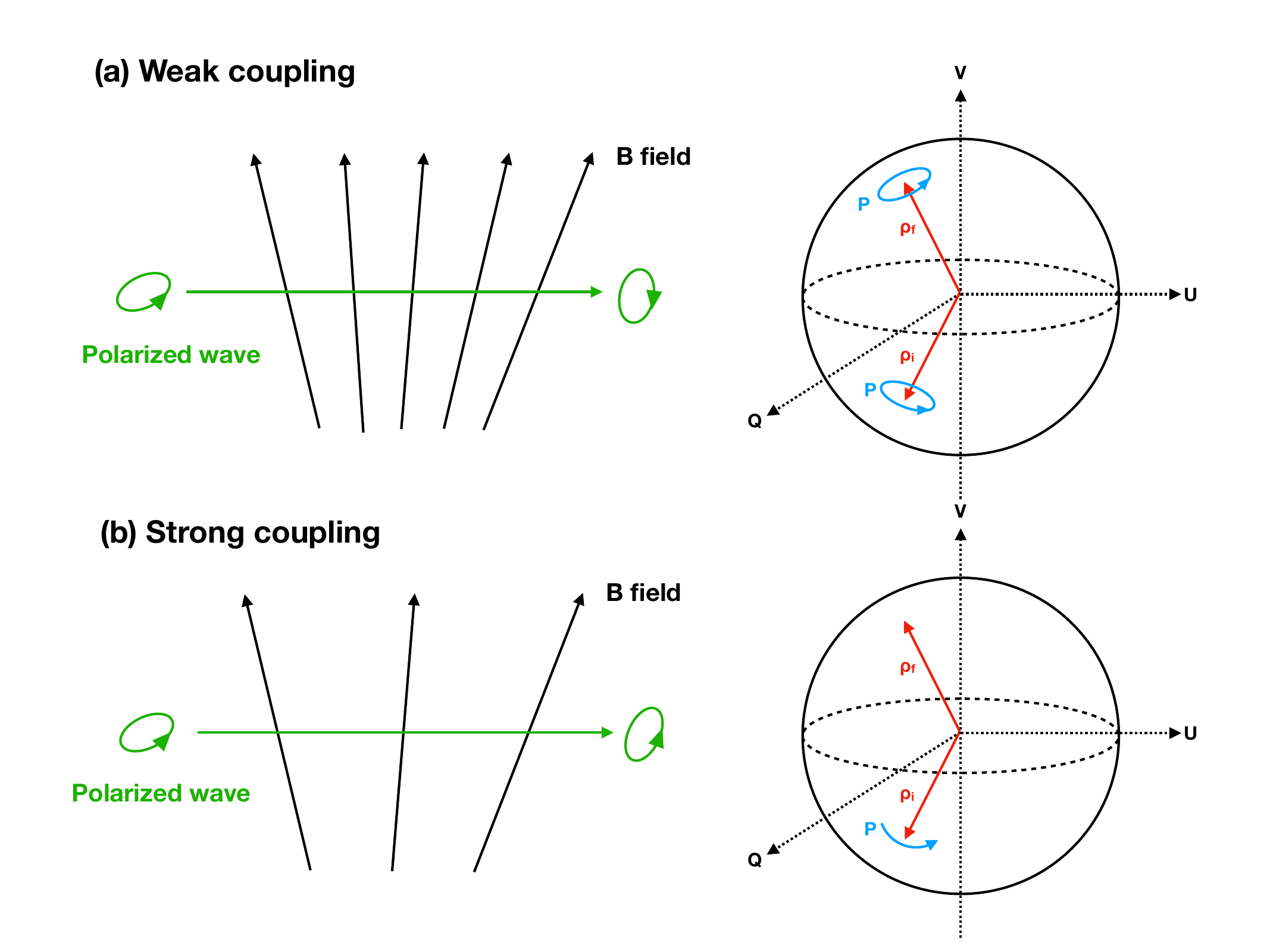} 
    \caption{Schematic of polarization evolution on the Poincar\'{e} sphere for Faraday conversion in the weak coupling ($C \ll 1$) and strong coupling ($C \gg 1$) regimes. In both panels, the natural mode axis, $\bm{\rho}$, evolves from an initial state $\bm{\rho}_i$ to a final state $\bm{\rho}_f$ as the wave traverses a quasi-transverse (QT) region. Panel (a): In the weak coupling limit, the wave's polarization vector, $\bm{P}$, remains locked to and is ``dragged'' by the evolving mode axis $\bm{\rho}$, resulting in a significant change to its final state. Panel (b): In the strong coupling limit, the polarization vector $\bm{P}$ cannot follow the rapid evolution of $\bm{\rho}$ and remains effectively ``frozen'' near its initial state.
    }\label{coupling} 
\end{figure} 

In a real scenario of the magnetosphere of a WD -- MD binary, the direction of the magnetic field along the line of sight can change gradually, i.e., the natural mode axes changes orientation continuously in the Poincar\'{e} sphere. We consider that the direction of the magnetic field suffers a sign reversal ($\theta\sim\pi/2$) at a point of $z=0$, so-called rQT region. 
Assuming that the coherent length of the magnetic field is $L_\theta$, then the polarization vector will rotate an angle of $\rho_LL_\theta$ around the natural mode axes in the Poincar\'{e} sphere, meanwhile, the natural mode axes will evolve an angle of $\delta\rho_V(L_\theta)/\rho_L$ at the same time, see Figure \ref{coupling}. Thus, one can define the coupling coefficient as \citep{Cohen60,Melrose10}
\be 
C\simeq\frac{\delta\rho_V(L_\theta)/\rho_L}{\rho_LL_\theta}\sim\frac{\rho_V}{\rho_L^2L_\theta}\simeq\fraction{\nu}{\nu_T}{4},
\ee 
where $\delta\rho_V\simeq\rho_V$ is adopted near the $QU$ plane, and the coupling frequency is 
\begin{align} 
\nu_T&=\frac{1}{2\pi}\fraction{\omega_p^2\omega_B^3L_\theta}{4c}{1/4}\simeq4.0~{\rm GHz}\fraction{n_e}{10^5~{\rm cm^{-3}}}{1/4}\nonumber\\ 
&\times\fraction{\mu_s}{10^{34}~{\rm G~cm^3}}{3/4}\fraction{P}{100~{\rm min}}{-4/3}\nonumber\\
&\times\fraction{M_s+M_c}{M_\odot}{-2/3}\fraction{L_\theta}{a}{-2}.\label{nuT}
\end{align} 
Notice that since the depolarization in the strong-field regime effectively erases most of the polarization information, we therefore focus primarily on the weak-field regime here.
The final polarization state is determined by the coupling strength, $C$. In the weak coupling regime ($C \ll 1$), the evolution is adiabatic. The rotation of the wave's polarization vector, $\bm{P}$, around the local natural mode axis, $\bm{\rho}$, is much faster than the evolution of the mode axis itself. As a result, the polarization vector remains ``locked'' to the natural mode axis. As $\bm{\rho}$ traverses the Poincar\'{e} sphere from one pole to the other through the QT region, it effectively ``drags'' the polarization vector $\bm{P}$ along with it, for instance, from the $-V$ hemisphere to the $+V$ hemisphere, or the opposite (see panel (a) of Figure \ref{coupling}). Such evolution can lead to dramatic observational consequences, including significant conversion between linear and circular polarization or even a reversal in the sign of the circular polarization component $V$.
Conversely, in the strong coupling regime ($C \gg 1$), the evolution is non-adiabatic. The natural mode axis, $\bm{\rho}$, now evolves much more rapidly than the polarization vector, $\bm{P}$, can rotate around it. Consequently, even as the mode axis $\bm{\rho}$ sweeps across the Poincar\'{e} sphere, the polarization vector $\bm{P}$ is left behind, remaining largely fixed near its initial position (see panel (b) of Figure \ref{coupling}). In this case, the wave's polarization is only minimally altered by its passage through the QT region, resulting in only a small change to its final polarization state.

For typical WD--MD binary parameters, the system generally resides in the weak coupling regime ($C \lesssim 1$), as the transition frequency, $\nu_T$, typically exceeds the observing frequency (Eq.(\ref{nuT})). In this limit, one might expect to observe a reversal in the sign of circular polarization ($V$) over a full orbital cycle if the initial state is circular, similar to the phenomenon recently reported for Terzan 5A by \citet{Li23}. Such a reversal would occur at superior conjunction, where the field strength along the line of sight is significantly larger than that at other orbital phases. However, for LPTs, this direct test is precluded by the strong beaming effect. The small observed pulse duty cycles (discussed in Section~\ref{beaming}) imply that we can only sample the polarization state from a narrow, fixed range of orbital phases, making a phase-dependent $V$-sign reversal unobservable.
Nevertheless, the weak coupling regime offers another powerful and testable prediction. The final polarization state is highly sensitive to the precise path the radio waves take through the magnetosphere. Due to the WD's rapid rotation relative to the orbital motion, the magnetic field geometry sampled by the line of sight will significantly vary from one orbit to the next. 
At different orbital periods, the degree of Faraday conversion would be different due to the modulation by the geometry of the magnetic field from the rotating WD. Thus, the weak coupling effect would cause an LPT to exhibit different polarization states at different periods, which can be tested by future observation. 

\section{Population of LPTs from WD -- M dwarf binaries}\label{population}

\subsection{Asynchronism of magnetic WD rotation}\label{asynchronism}

A unifying requirement for both the unipolar-inductor (Section~\ref{stage2}) and the magnetosphere-interaction (Section~\ref{stage3}) mechanisms is the presence of asynchronism ($\Delta\Omega \neq 0$). In the former, it is the relative motion between the WD's magnetosphere and the MD's conductive body that drives the current. In the latter, it is the differential rotation between the two magnetospheres that powers reconnection. Therefore, we conclude that a state of asynchronous rotation is a fundamental prerequisite for producing the coherent radio emission of LPTs in WD--MD binaries.
Furthermore, the prevalence of this asynchronous state has profound implications for the LPT population. The evolutionary timescale over which a binary remains asynchronous will directly determine the fraction of WD--MD systems that are observable as LPTs.

\begin{figure} 
    \centering
    \includegraphics[width = 0.9\linewidth, trim = 0 0 0 0, clip]{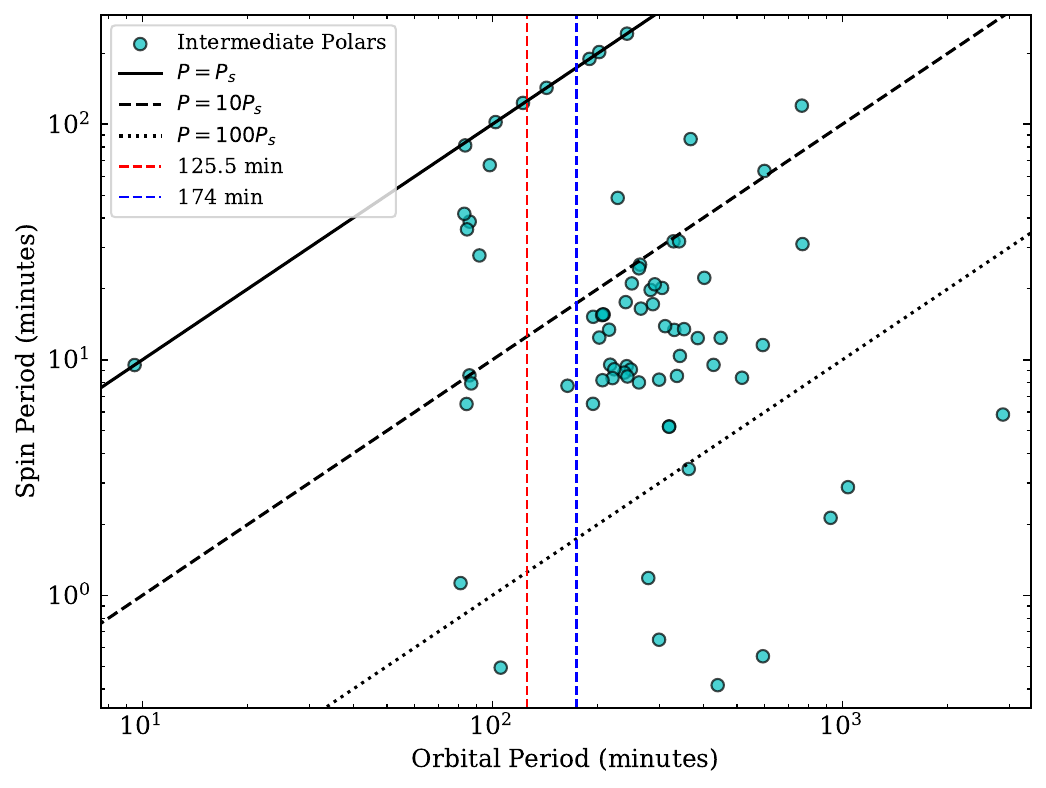} 
    \caption{Orbital period versus WD spin period for the known population of Intermediate Polars (IPs). The data of IPs is from Koji Mukai's ``The Intermediate Polars'' homepage (\url{https://asd.gsfc.nasa.gov/Koji.Mukai/iphome/iphome.html}, also see \citet{Mukai17}). The dashed red and blue lines denote the periods of 125.5 min and 174 min of ILT J1101+5521 and GLEAM-X J0704-37 that have been confirmed from the magnetic WD -- MD binaries.}\label{spinorb} 
\end{figure}

The degree of asynchronism in mCVs provides a crucial context for understanding LPTs. The mCV population is sharply divided into two subclasses based on this property. In Polars, the powerful magnetic fields of the WDs with magnetic moment of $\mu_s \gtrsim 5\times10^{33}{\rm G~cm^3}$ are sufficient to enforce spin-orbit synchronization, locking their spin periods to the orbital periods. In contrast, IPs possess weaker magnetic moments that are insufficient to achieve such phase-locking. Consequently, their spin periods are systematically shorter than their orbital periods.
The distribution of IPs in the orbital period vs. spin period diagram is revealing (see Figure \ref{spinorb}). While a small fraction of IPs lie near the synchronous line alongside Polars, the majority cluster in a region defined by $P_s \sim (10^{-2}$--$10^{-1}) P_{\rm orb}$ \citep{Barrett88, Wu91}. This corresponds to a state of extreme asynchronicity, with $\Delta\Omega/\Omega \gg 1$. The requirement of strong asynchronism for LPT production thus points to a compelling analogy: LPT progenitors share an asynchronous characteristic with the IP population but in a pre-accretion, detached evolutionary state.

The observed period of two LPTs (ILT J1101+5521 and GLEAM-X J0704-37) has been confirmed to correspond to the orbital period,
\be 
P_{\rm obs}=P.
\ee 
Due to the beaming effect of the radiation, although the intrinsic radiation is persistent in all phases, the beaming effect causes it to only be seen at specific phases.
On the other hand, due to the anisotropy of magnetic field distribution of the WD's magnetosphere and the asynchronism of the WD rotation, both the intensity and polarization of the radiation will be modulated by the beat period of $P_{\rm beat}$ of 
\be 
P_{\rm beat}=\left|\frac{1}{1/P_s-1/P}\right|.
\ee
For a synchronous Polar, phase-locking ($P_s \simeq P$) results in a very long beat period, rendering the modulation unobservable. For a typical asynchronous IP, however, the beat period is approximately $P_{\rm beat} \sim P_s$, which is significantly shorter than the orbital period $P$. The detection of such a beat signal, manifested as a periodic evolution in the peak flux or polarization state of an LPT, would provide a powerful test of this model in future observations. 
Furthermore, if the LPT flux is simultaneously influenced by the beat effect and other processes (e.g., intrinsic instability variations), isolating the beat modulation component becomes considerably more challenging. In such cases, instead of simply monitoring the flux changes of individual pulses, a statistical analysis is required to determine whether beat modulation is indeed present.

The fraction of WD--MD binaries observable as LPTs is ultimately determined by the duration over which they can maintain an asynchronous state. This asynchronism timescale is governed by the braking torque that acts to synchronize the WD's spin with the orbit. In the physical scenarios we have considered, this braking torque arises from the magnetic interaction between the WD's magnetosphere and either the MD's conductive body or its own magnetosphere \citep{Joss79, Lamb83, Campbell85}.
The magnitude of this synchronizing torque, $N_B$, is directly related to the energy dissipation rate, $\dot{E}$, by $N_B \sim \dot{E}/\Delta\Omega$. This allows us to estimate the synchronization timescale, $\tau_{\rm syn}$, which is the characteristic time required for the torque to eliminate the differential rotation: $\tau_{\rm syn} \sim I_s \Delta\Omega / N_B$, where $I_s = (2/5)M_s R_s^2$ is the moment of inertia of the WD. By substituting the energy dissipation rate for the non-saturated unipolar-inductor mechanism from Eq.(\ref{diss_up}), we can derive the corresponding synchronization timescale as
\begin{align} 
\tau_{\rm syn}&\simeq5.7\times10^8~{\rm yr}\fractionz{M_s}{0.8M_\odot}\fraction{R_s}{10^9~{\rm cm}}{2}\nonumber\\ 
&\times\fraction{\mu_s}{10^{34}~{\rm G~cm^3}}{-2}\fraction{R_c}{10^{10}~{\rm cm}}{-2}\fraction{P}{100~{\rm min}}{8/3}\nonumber\\ 
&\times\fraction{M_s+M_c}{M_\odot}{4/3}.
\end{align} 
Similarity, according to Eq.(\ref{diss_up2}), for the saturated unipolar-inductor mechanism, synchronization timescale is 
\begin{align} 
\tau_{\rm syn}&\simeq1.2\times10^6~{\rm yr}\fractionz{M_s}{0.8M_\odot}\fraction{R_s}{10^9~{\rm cm}}{2}\nonumber\\ 
&\times\fraction{\mu_s}{10^{34}~{\rm G~cm^3}}{-2}\fraction{R_c}{10^{10}~{\rm cm}}{-2}\fraction{P}{100~{\rm min}}{7/3}\nonumber\\ 
&\times\fraction{M_s+M_c}{M_\odot}{5/3}\fractionz{\Delta\Omega}{\Omega}.
\end{align}
For the magnetosphere-interaction process, according to Eq.(\ref{diss_mc}), the synchronism timescale is 
\begin{align} 
\tau_{\rm syn}&\simeq2.6\times10^5~{\rm yr}\fractionz{M_s}{0.8M_\odot}\fraction{R_s}{10^9~{\rm cm}}{2}\nonumber\\ 
&\times\fraction{\mu_s}{10^{34}~{\rm G~cm^3}}{-1}\fraction{\mu_c}{10^{33}~{\rm G~cm^3}}{-1}\fractionz{P}{100~{\rm min}}\nonumber\\ 
&\times\fractionz{M_s+M_c}{M_\odot}\fractionz{\Delta\Omega}{\Omega}.
\end{align}

To estimate the fraction of LPT-hosting systems within the mCV population, we need to compare the timescale over which a system remains an LPT (i.e., the asynchronism timescale, $\tau_{\rm syn}$) with the total evolutionary timescale of a typical mCV. The long-term evolution of mCVs is primarily driven by the loss of orbital angular momentum, a process dominated by magnetic braking or gravitational radiation.
For the WD--MD binaries under consideration, both magnetic braking and gravitational radiation might be important drivers of their evolution. A generalized law describing the rate of angular momentum loss due to magnetic braking, $\dot{J}_{\rm MB}$, for a binary with total orbital angular momentum $J_{\rm orb}$ is given by \citet{Rappaport83} as
\be 
\frac{\dot J_{\rm MB}}{J_{\rm orb}}=-3.8\times10^{-30}~{\rm s^{-1}}f\frac{R_\odot^4(R_c/R_\odot)^\gamma GM^2}{a^5M_s},\label{rap83}
\ee
where $\gamma$ is the magnetic braking index expressing the dependence of the braking on the MD radius, $f$ is a constant of order unity, $M=M_s+M_c$ is the total mass. 
Using the mass-radius relation of Eq.(\ref{RM_relation}) and taking $\gamma\simeq4$ as a fiducial value in its possible range, the timescale of the magnetic braking $\tau_{\rm MB,0}$ is given by 
\begin{align} 
\tau_{\rm MB,0}&\sim\frac{J_{\rm orb}}{\dot J_{\rm MB}}\simeq1.7\times10^8~{\rm yr}f^{-1}\fraction{M_s+M_c}{M_\odot}{-1/3}\nonumber\\ 
&\times\fractionz{M_s}{0.8M_\odot}\fraction{M_c}{0.2M_\odot}{-3.2}\fraction{P}{100~{\rm min}}{10/3}.
\end{align}
Due to the strong magnetic field of the WD partially suppressing magnetic braking, mCVs do not experience an abrupt halt in accretion in the period gap of $2-3$ hr, unlike non-magnetic CVs. This results in a longer braking timescale, $\tau_{\rm MB} \sim \xi_{\Phi} \tau_{\rm MB,0}$, where $\xi_{\Phi} \gtrsim 1$ is a dimensionless parameter that depends on the fraction of open field lines, $\Phi$, in the magnetosphere \citep{Li94,Belloni20}.
On the other hand, the timescale of the gravitational radiation is \citep{Landau75}
\begin{align} 
\tau_{\rm GW}&\sim\frac{J_{\rm orb}}{\dot J_{\rm GW}}=\frac{5c^5}{32G^3}\frac{a^4}{M_sM_cM}\simeq1.9\times10^9~{\rm yr}\nonumber\\ 
&\times\fraction{M_s+M_c}{M_\odot}{1/3}\fraction{M_s}{0.8M_\odot}{-1}\fraction{M_c}{0.2M_\odot}{-1}\nonumber\\
&\times\fraction{P}{100~{\rm min}}{8/3}.\label{JGW}
\end{align} 
It is important to recognize that for a mCV in long-term evolution, mass transfer via Roche-lobe overflow enforces a specific relationship between the MD's mass, radius, and the binary's orbital period. By combining Kepler's third law (Eq.(\ref{Kepler3})) with the mass-radius relation (Eq.(\ref{RM_relation})) and the Roche-lobe scale (Eq.(\ref{roche})), one can derive the scaling relations $R_c \propto M_c^{0.8}$ and $M_c \propto P^{1.43}$ for the case where $M_c \ll M_s$.
Substituting this period-dependent mass into our expressions for the evolutionary timescales yields their own dependence on the orbital period. We find that the magnetic braking timescale scales as $\tau_{\rm MB,0} \propto P^{-1.24}$, while the gravitational radiation timescale scales as $\tau_{\rm GW} \propto P^{1.24}$. This confirms the standard picture of WD--MD binary evolution: magnetic braking is the dominant angular momentum loss mechanism at long orbital periods, whereas gravitational radiation dominates at short periods. With these evolutionary timescales established, the number ratio of the LPT population to the total mCV population can now be estimated as
\be 
f_{\rm LPT}\sim\frac{\tau_{\rm syn}}{\min(\tau_{\rm MB},\tau_{\rm GW})}.\label{flpt}
\ee 
This ratio implicitly assume that all mCVs undergo an asynchronous LPT phase. For relatively young binary systems whose evolutionary timescales are shorter than the typical duration of the LPT phase, their contribution to the observed LPT fraction should be low. For these reasons, the estimated $f_{\rm LPT}$ is not expected to be significantly overestimated.

Evaluating these ratios for the typical system parameters yields distinct predictions for the two models. For the unipolar-inductor mechanism, we predict a relatively large LPT population fraction, $f_{\rm LPT} \lesssim (10^{-3}-1$). In contrast, the magnetosphere-interaction model predicts a much smaller fraction, $f_{\rm LPT} \lesssim (10^{-4}-10^{-3})$.
These contrasting predictions highlight a fundamental trade-off between the two scenarios. The unipolar-inductor mechanism, despite its lower energy dissipation rate (especially in the unsaturated regime), can sustain an asynchronous state for a relatively long time, thus predicting that LPTs should be relatively common. Conversely, the magnetosphere-interaction process can be more powerful, capable of producing higher luminosities for the same radiation efficiency. However, this powerful interaction also leads to rapid synchronization, dramatically shortening the LPT lifetime and thus predicting a relatively rarer population. Distinguishing between these two predictions (i.e., a large population of lower-luminosity LPTs versus a small population of higher luminosity LPTs) presents a key observational test for future population studies.

\subsection{Period distribution}

For the two sources (ILT J1101+5521 and GLEAM-X J0704-37) identified as WD--MD binaries, this periodicity has been confirmed to be orbital in nature. This direct link between the LPT phenomenon and binary orbital evolution provides a powerful framework for making further predictions. Next, we will leverage this connection to predict the period distribution of the LPT population originating from magnetic WD--MD binaries.

For a binary system with a circular orbit, the orbital angular momentum is 
\be 
J_{\rm orb}=\fraction{Ga}{M}{1/2}M_sM_c\propto P^{1/3},\label{Jorb}
\ee 
where $M=M_s+M_c$ is the total mass of the binary system. 
The orbital period evolution is due to angular momentum loss and its derivative is 
\begin{align} 
\dot P=6\pi\fraction{J_{\rm orb}}{GM}{2}\fraction{M}{M_sM_c}{3}\dot J_{\rm orb}\propto \dot J_{\rm orb}P^{2/3}.\label{dP}
\end{align}
To derive the period distribution, we first establish the evolutionary behavior of the binary in the pre-mCV phase. Since LPTs are, by our hypothesis, not undergoing Roche-lobe overflow, we can assume that the mass transfer rate is negligible ($\dot{M}_s \sim \dot{M}_c \sim 0$). This implies that the MD's mass is independent of the orbital period during this evolutionary stage. Under this assumption, the evolution of the orbital period is dictated solely by angular momentum loss.
We consider the two dominant mechanisms for angular momentum loss. For magnetic braking, combining Eq.(\ref{rap83}), Eq.(\ref{Jorb}) and Eq.(\ref{dP}) yields a period evolution rate of $\dot{P} \propto P^{-7/3}$. For gravitational radiation, a similar combination of Eq.(\ref{JGW}), Eq.(\ref{Jorb}) and Eq.(\ref{dP}) gives $\dot{P} \propto P^{-5/3}$. The characteristic lifetime of an LPT at a given period is then $\tau \propto |P/\dot{P}|$. Assuming a constant formation rate for LPT progenitor systems, the number of systems at a given age, $dN = f_\tau(\tau)d\tau$, should be constant over time. By transforming variables from age to period ($dN = f_P(P)dP = f_\tau(\tau)d\tau$), we can derive the intrinsic period distribution as
\be 
f_P(P)\propto \frac{d\tau}{d P}\propto \frac{d|P/\dot P|}{d P}\propto P^{\alpha},
\ee 
where $\alpha$ is defined by $\dot P\propto P^{-\alpha}$, $\alpha=7/3$ and $5/3$ correspond to the magnetic braking and gravitational radiation, respectively.

\subsection{Luminosity function}\label{LF} 

\begin{table*}[htbp]
\centering
\caption{Predicted period distribution $f_P(P)$ and luminosity function $f_L(L)$ of LPTs from the magnetic WD -- MD binaries for different possible scenarios}\label{table}
\begin{tabular}{llcc}
\toprule
& & Magnetic Braking & Gravitational Radiation \\
\midrule
\multicolumn{2}{l}{Period Distribution} & $P^{2.33}$ & $P^{1.67}$ \\
\midrule
\multirow{3}{*}{Luminosity Function} 
 & Unipolar-inductor (Unsaturated) & $L^{-2.25}$ & $L^{-2.00}$ \\
 & Unipolar-inductor (Saturated) & $L^{-2.00}$ & $L^{-1.80}$ \\
 & Magnetosphere-interaction & $L^{-2.67}$ & $L^{-2.33}$ \\ 
\bottomrule
\end{tabular}
\\
\vspace{0.3em}
\footnotesize Note: For the unipolar-inductor mechanism, the period range is $P_{\rm acc}\lesssim P\lesssim P_{\rm UI}$ given by Eq.(\ref{Pui}); For magnetosphere-interaction mechanism, the period range is $P\gtrsim P_{\rm UI}$ given by Eq.(\ref{Pmi}). 
\end{table*}

According to Eq.(\ref{diss_up}), Eq.(\ref{diss_up2}), and Eq.(\ref{diss_mc}), both the energy dissipation rates of the unipolar-inductor and magnetosphere-interaction mechanisms have a strong dependence of the orbital period. We consider a significant asynchronism with $P\gg P_s$. For the unsaturated unipolar-inductor mechanism with $1\ll\Delta\Omega/\Omega\ll(\Delta\Omega/\Omega)_{\rm cr}$, one has $\dot E\propto P^{-14/3}(\Delta\Omega/\Omega)^2\propto P^{-8/3}$; For the saturated unipolar-inductor mechanism with $\Delta\Omega/\Omega\gtrsim(\Delta\Omega/\Omega)_{\rm cr}$, one has $\dot E\propto P^{-13/3}(\Delta\Omega/\Omega)\propto P^{-10/3}$.
For magnetic reconnection, the energy dissipation rate has a period-dependence of $\dot E\propto P^{-3}(\Delta\Omega/\Omega)\propto P^{-2}$. 
We define an index of $\beta$ with $\dot E\propto P^{-\beta}$. $\beta=8/3, 10/3$ and $2$ correspond to unsaturated unipolar-inductor mechanism, saturated unipolar-inductor mechanism, and magnetosphere interaction, respectively. 
Given that this strong period dependence is the dominant factor determining the dissipation rate compared to other system parameters, we can use it to predict the LPT luminosity function. Assuming that the orbital period distribution is $f_P(P)$ and the LPT luminosity is proportional to the energy dissipation rate, $L\propto\dot E\propto P^{-\beta}$, then the luminosity function $f_L(L)$ is given by $f_L(L)dL=f_P(P)dP$, leading to 
\begin{align} 
f_L(L)&=\frac{P}{L}\fractionz{d\ln P}{d\ln L}f_P(P)\nonumber\\ 
&\propto L^{-\frac{\beta+1}{\beta}}f_P[P(L)]\propto L^{-\frac{\alpha+\beta+1}{\beta}}.
\end{align} 
Thus, the luminosity function is predicted to be $f_L(L)\propto L^{-(1.80-2.67)}$ for the combination of different scenarios, as shown in Table \ref{table}.

\subsection{Volumetric number density}\label{density}

The volumetric number density of LPTs originating from WD--MD systems can be statistically constrained by the distance to the nearest detected source. Following a Bayesian approach similar to that used for constraining the population of fast radio bursts \citep{Lu22}, we can first estimate the observable number density, $n_{*,\rm obs}$, based on the volume, $V_1$, enclosed by the closest known LPT.
It is crucial to distinguish this observable density from the intrinsic density, $n_*$. The beaming of the LPT emission means that we only detect sources whose emission cones happen to sweep across our line of sight. To account for the much larger population of sources permanently beamed away from us, we must apply a beaming correction. 
We model the LPT emission as a conical beam with a half-opening angle of $\Theta_{\rm obs}$. As this beam co-rotates with the binary system, it sweeps out a solid angle of approximately $2\pi \Theta_{\rm obs}$ ($\Theta_{\rm obs}\ll1$) over one orbital period. The probability of a randomly oriented system being observable is the ratio of this swept solid angle to the full sky ($4\pi$), which defines the beaming factor $f_b = 2\pi\Theta_{\rm obs}/ 4\pi = \Theta_{\rm obs}/2$. The intrinsic number density, $n_*$, is therefore related to the observable density, $n_{*,\rm obs}$, by correcting for this geometric selection effect: $n_* = n_{*,\rm obs} / f_b = n_{*,\rm obs} (2/\Theta_{\rm obs})$.

The probability distribution for the observable number density is given by the exponential distribution $dP/dn_{*,\rm obs} = V_1 \exp(-V_1 n_{*,\rm obs})$. 
The corresponding cumulative distribution is $P(<n_{*,\rm obs}) = 1 - \exp(-V_1 n_{*,\rm obs})$. From this, one can derive the 68\% confidence interval for $n_{*,\rm obs}$ as \citep{Lu22}: 
\be
n_{*,\rm obs}\in(0.17,1.8)V_1^{-1}. 
\ee 
At present, the closest LPT identified as a WD -- MD binary, GLEAM-X J0704-37, has a distance of 380 pc \citep{Rodriguez25}. Besides, CHIME J0630+25 at 170 pc is the closest LPT discovered to date, although its origin is still not confirmed \citep{Dong25}.
Thus, the volumetric number density is estimated to be $n_{*}=n_{*,\rm obs}(2/\Theta_{\rm obs})$,
\begin{align}
n_{*}\in 
\begin{cases} 
(0.15-1.6)\times10^{-7}~{\rm pc^{-3}}(\Theta_{\rm obs}/0.1)^{-1}, &\text{J0704-37} \\
(1.6-17)\times10^{-7}~{\rm pc^{-3}}(\Theta_{\rm obs}/0.1)^{-1}, &\text{J0630+25}
\end{cases}
\end{align}
The two number densities derived above are based on the distances to the nearest LPTs, GLEAM-X J0704-37 and CHIME J0630+25, respectively. These calculations yield an estimated intrinsic volumetric number density for LPTs from WD--MD binaries of $n_* \sim (10^{-8}-10^{-6})\,(\Theta_{\rm obs}/0.1)^{-1}\,\text{pc}^{-3}$.
We can now compare this estimated LPT density to the known densities of related populations. The total space density of CVs and mCVs, including a large undiscovered population, is estimated to be $n_{*,\text{CV}} \sim (0.5-2) \times 10^{-4}\,\text{pc}^{-3}$ \citep{deKool92}. Thus, the LPT number density is several orders of magnitude smaller than the total CV population.
The comparison with the mCV population, however, is far more revealing. The observed space densities of Polars and IPs are $n_{*,\text{pol}} \sim 3 \times 10^{-7}\,\text{pc}^{-3}$ \citep{Warner95} and $n_{*,\text{ip}} \sim (10^{-7}-10^{-6})\,\text{pc}^{-3}$ \citep{Pretorius14}, respectively. 
The derived number density for LPTs is not much less than the density of known mCVs, although it largely depends on the uncertain beaming factor. 
From this perspective, it is plausible that LPTs predominantly originate from the unipolar-inductor mechanism, though a smaller contribution from the magnetosphere-interaction mechanism (i.e., magnetic reconnection) cannot be ruled out, see Section \ref{asynchronism} and Eq.(\ref{flpt}).
In conclusion, LPTs originating from WD--MD binaries are not a distinct class of objects, but are instead a manifestation of the mCV population itself, likely representing a significant fraction of these systems observed during their transient, pre-accretion evolutionary phase.

\subsection{Cooling issue of magnetic WDs}\label{WDcooling} 

It is noteworthy that the WDs in both ILT J1101+5521 ($T_{\rm eff} = (4500–7500)~{\rm K}$; \citealt{deRuiter25}) and GLEAM-X J0704–37 ($T_{\rm eff} = 7300~{\rm K}$; \citealt{Rodriguez25}) exhibit very low effective temperatures, providing important constraints on the evolutionary history of magnetic WDs (also see the comment in \citealt{CastroSegura25}). 
For an isolated WD, its luminosity $L$ decreases with age $t$ following the Mestel cooling law, $L \propto t^{-7/5}$, which can be expressed as (see Section 37.3 of \citet{Kippenhahn13}) 
\be 
L(t)=2.8\times10^{-4}L_\odot\fractionz{M_s}{0.8M_\odot}\fraction{t}{1~{\rm Gyr}}{-7/5},
\ee
with the mass number per ion of $A=14$ for CO-WDs. Using the mass-radius relation of WDs, $R_s\sim 0.01R_\odot (M_s/M_\odot)^{-1/3}$ for $M_s\lesssim M_{\rm ch}\simeq1.4M_\odot$, the evolution of the effective temperature can be written as
\be 
T_{\rm eff}(t)=7200~{\rm K}\fraction{M_s}{0.8M_\odot}{5/12}\fraction{t}{1~{\rm Gyr}}{-7/20}.
\ee 
Therefore, if the WD magnetic field in LPT sources arises from the fossil field at birth \citep{Braithwaite04} or the dynamo action during common-envelope evolution \citep{Tout08}, the negligible accretion in the pre-LPT phase results in WD cooling behavior consistent with that of an isolated WD and the WD age is constrained to several Gyr. 

On the other hand, if the WD's field is generated by a crystallization- and rotation-driven dynamo during the evolution from a non-magnetic CV to a mCV \citep{Schreiber21}, the effective temperature of the magnetic WD in the LPT sources would be higher due to the compressional heating by the accretion in the preceding CV phase and the relatively short cooling timescale of the pre-mCV phase, also see Figure \ref{accrate}. The bolometric luminosity of the WD at the beginning of the pre-mCV phase can reach $T_{\rm eff,0}\sim\text{a few}\times10^4~{\rm K}$ for a WD mass of $M_s\sim(0.6-1.2)M_\odot$ and an accretion rate of $\dot M\sim (10^{-11}-10^{-8})M_\odot{\rm yr^{-1}}$ (see Figure 10 of \citet{Townsley04}), At time $t$ after the beginning of the pre-mCV phase, the effective temperature of the WD can be estimated as
\be 
T_{\rm eff}(t)\simeq T_{\rm eff,0}\min\left[1,\fraction{t}{t_0}{-7/20}\right],
\ee 
where $t_0 \sim \tau_{\rm cool} \sim \text{a few}\times(0.1-1)~{\rm Gyr}$ represents the typical cooling timescale of a WD for an accretion rate of $\dot M \sim (10^{-11}-10^{-8})~M_\odot~{\rm yr^{-1}}$, maintained near its equilibrium value \citep{Townsley04}. 
As shown in Figure 1 of \citet{Schreiber21}, the pre-mCV phase in this scenario lasts only tens of millions of years, much shorter than the typical WD cooling timescale. Consequently, this evolutionary scenario alone cannot account for the observed effective temperatures of WDs in LPT sources, indicating that a special cooling mechanism or a longer pre-mCV timescale is required in this picture.

\section{Discussions and Conclusion}\label{discussion}

LPTs are a newly identified class of coherent radio sources with periods spanning from minutes to hours. The recent confirmation that two such sources, ILT J1101+5521 and GLEAM-X J0704-37, originate from WD--MD binary systems has provided a critical breakthrough in understanding their nature \citep{deRuiter25, Hurley-Walker24, Rodriguez25}.
In this study, we proposed a model wherein at least a subset of the LPT population arises from magnetic WD--MD binaries during their detached, pre-mCV evolutionary phase, prior to the onset of Roche-lobe overflow. The partial overlap between the observed period distribution of LPTs and the orbital periods of CVs and mCVs (Figure \ref{CVperiod}) lends strong support to this scenario and suggests that many more LPTs await discovery in such binary systems. Identifying these counterparts is observationally feasible. 
For a typical M3--M5 dwarf companion, as found in ILT J1101+5521 and GLEAM-X J0704-37, its absolute magnitude is $M_V \simeq (11-14)$ mag. For a telescope with a limiting magnitude of $m_{\rm th} = 23$ mag, such a companion would be detectable within a distance of $d = 10^{(m_{\rm th}-M_V+5)/5} \simeq (0.6-2.5)$ kpc, a range that comfortably encompasses the distances of currently known LPTs.
Therefore, a dedicated observational campaign to systematically monitor the optical counterparts of known LPTs would be highly valuable. Such a program could not only confirm the binary nature of more sources but also place crucial constraints on their evolutionary state and physical parameters, providing a powerful test of the models presented in this work. 

Notably, the observed period distribution of LPTs presents a significant challenge to a single origin. Approximately half of the known LPTs exhibit periods substantially shorter than the canonical $\sim$80-minute orbital period minimum for CVs and mCVs. 
This short-period population cannot be readily explained by standard binary evolution models, which predict that as an MD companion loses mass and becomes degenerate, its radius begins to increase, causing the binary orbit to widen again by enhanced accretion, i.e., the well-known ``period bounce'' phenomenon \citep{Paczynski81, Rappaport83}. 
The existence of this sub-80-minute population therefore strongly suggests a distinct physical origin. It is highly plausible that these shorter-period LPTs are not powered by orbital motion at all, but rather by the spin of compact objects, such as neutron stars or WDs. This hypothesis is bolstered by independent observational evidence, including the association of ASKAP/DART-J1832 with a supernova remnant and the long-period X-ray pulsations from the magnetar candidate IE 161348-5055 \citep{DeLuca06, Li24, Wang25}. 

It is interesting that the orbital periods of ILT J1101+5521 and GLEAM-X J0704-37 place them squarely within the so-called ``period gap'' ($P \simeq (2-3)$ hr) of the non-magnetic CV population (see Figure \ref{CVperiod}). This gap is thought to represent a phase in which standard magnetic braking ceases, temporarily halting Roche-lobe overflow and making these systems effectively unobservable via accretion-powered emission.
While this period gap is a prominent feature in the non-magnetic CV distribution, it is significantly less pronounced for mCVs. It is theorized that the strong magnetic fields in some mCVs can sustain a high rate of angular momentum loss, thereby maintaining mass transfer even through this period range. Nevertheless, a detached, non-accreting population of magnetic binaries is still expected to exist within the gap. As we have demonstrated in Section \ref{stage}, there is a substantial parameter space in which a magnetic WD--MD binary can reside in a pre-mCV, detached state.

In this study, we have argued that LPTs originating from WD--MD binaries serve as crucial probes of the pre-mCV evolutionary phase. A key conclusion from our analysis is that the intense, coherent radio emission characteristic of LPTs is fundamentally incompatible with a high-density plasma environment. This effectively rules out standard accretion-dominated scenarios ($\dot{M} \gtrsim10^{-10}M_\odot{\rm yr^{-1}}$) as the origin of these sources.
Instead, we find that the energy requirements and observational properties of LPTs can be successfully explained by two distinct mechanisms operating in a detached, non-accreting binary: the unipolar-inductor mechanism and the magnetosphere interaction. The dominant mechanism in any given system is mainly determined by the relative magnetic moments of the WD and MD. Crucially, both models naturally produce a geometrically-beamed radiation pattern, consistent with the small pulse duty cycles ($\Delta t/P \sim (10^{-3}-10^{-1}$) observed in LPTs. The precise beaming angle in both scenarios is governed by the large-scale magnetic field geometry of the WD.

Due to the different dissipation regions and magnetospheric configuration in the unipolar-inductor phase and the magnetosphere-interaction phase, some observable features are expected to differ, providing potential diagnostics for identifying the dominant dissipation channel. First, since the magnetic field geometry in unipolar-inductor phase is much simpler than that in the magnetosphere-interaction phase, a polarization evolution pattern modulated by the beat period is more likely to occur during the unipolar-inductor phase. 
Second, from a statistical perspective, these two mechanisms are expected to exhibit different luminosity functions (see Section \ref{LF}). However, because LPTs are likely to consist of two distinct populations, such a statistical verification requires a separate classification of WD–MD binaries. Third, under the unipolar-inductor mechanism, the interaction between the WD’s magnetic field and the MD (or its wind near the surface) may produce non-thermal emission from radio to optical wavelengths, as observed in AR Scorpii, particularly when the WD's spin is sufficiently fast to trigger pair production \citep{Marsh16,Geng16}. 

Asynchronism ($\Delta\Omega \neq 0$) is a necessary condition for LPT activity. Both the unipolar-inductor mechanism and the magnetosphere interaction are powered by the differential rotation between the WD's magnetosphere and the MD or its own magnetosphere. This requirement provides a powerful diagnostic tool for identifying the nature of LPT progenitors. From this perspective, LPTs are not analogous to synchronous Polars, but rather to the asynchronous IP population, which is characterized by rapid stellar rotation relative to the orbit ($P_s \sim (10^{-2}-10^{-1}) P_{\rm orb}$). 
This state of asynchronous rotation, combined with the anisotropic nature of the WD's magnetosphere, leads to a distinct and testable observational prediction. We predict that both the intensity and the polarization of the LPT emission should exhibit a modulation at the beat period between the spin and orbital frequencies. 
Moreover, if the LPT flux is affected by both beat modulation and other processes, statistical analysis rather than individual pulse monitoring is required to identify the presence of beat modulation. 

Regarding the radiation mechanism itself, we find that the LCDM, a specific form of ECME, provides a compelling explanation for the observed properties of LPTs. This mechanism arises naturally within the binary's magnetosphere, as the converging magnetic field lines ($\nabla\bm{B} \parallel \bm{B}$) inherently form the magnetic mirrors required for its operation. The continuous precipitation of small-pitch-angle electrons through these mirrors naturally establishes and maintains the anisotropic, loss-cone distribution necessary to drive the maser.
This model makes several key, testable predictions. First, it can naturally produce the high brightness temperatures observed in LPTs, with predicted values reaching up to $T_B \sim 10^{13}$ K. Second, the maser is expected to operate in a self-regulating, saturated state, flickering sporadically around its emission threshold. This predicts the existence of fine-scale temporal substructures within the broader LPT pulse profile. The brightness temperature of these individual sub-bursts could be significantly higher than the time-averaged value, reflecting moments of saturated, peak emission.

Once generated, the radiation must traverse the binary's magnetosphere, where its polarization state can be profoundly altered by propagation effects. 
Along certain lines of sight, particularly those passing close to either the WD or the MD, the local electron cyclotron frequency may exceed the wave frequency, resulting in significant depolarization. In other directions, where the radiation path deviates from the WD/MD, the conversion frequency, $\nu_{\rm FC}$, can be several times the observing frequency, resulting in a distinct, frequency-dependent polarization signature observable across a typical radio telescope bandwidth.
Furthermore, our analysis indicates that for typical WD--MD binary systems, the propagation is generally in the weak coupling regime. The final polarization state is highly sensitive to the exact path taken through the magnetosphere. Due to the WD's rapid and asynchronous rotation, the magnetic field geometry sampled by our line of sight varies from one orbit to the next. The weak coupling effect translates these magnetic field variations into measurable changes in the final polarization state. We predict that the polarization state can be significantly different at different orbital periods.

Finally, we have made key statistical predictions for the LPT population originating from WD--MD binaries. We derive an intrinsic period distribution of $f_P(P) \propto P^{(1.67-2.33)}$ and a corresponding luminosity function of $f_L(L) \propto L^{-(1.80-2.67)}$. These power-law distributions provide a direct and falsifiable prediction that can be rigorously tested by future observations as the sample of known LPTs grows. 
At last, since the generation of LPTs requires that the WD should be magnetic, the observed low effective temperatures of ILT J1101+5521 and GLEAM-X J0704-37 suggest that the magnetic field of the WDs in these LPT sources more likely originate from the fossil field at birth \citep{Braithwaite04} or the dynamo action during common-envelope evolution \citep{Tout08}, rather than a crystallization- and rotation-driven dynamo during the evolution from a non-magnetic CV to a fully magnetic CV \citep{Schreiber21}. This feature provides crucial clues for investigating the formation channels of magnetic WDs.

\section*{Acknowledges}
We thank the anonymous referee for providing helpful comments and suggestions.
We also acknowledge the helpful discussions with Ingrid Pelisoli, Weiwei Zhu, Noel Castro Segura, Biping Gong, Jieshuang Wang, Ye Li, and Boyang Liu.
This work made use of data of PolarCat \url{https://www.aip.de/en/members/axel-schwope/polarcat/}, the data of IPs is from Koji Mukai's ``The Intermediate Polars'' homepage (\url{https://asd.gsfc.nasa.gov/Koji.Mukai/iphome/iphome.html}, also see \citet{Mukai17}), and the data of CVs is from SDSS I to IV collated from multiple archival data sets \citep{Inight23}. 
This work is supported by the National Natural Science Foundation of China (No. 12473047), the National Key Research and Development Program of China (No. 2024YFA1611603) and the Yunnan Key Laboratory of Survey Science (No. 202449CE340002).


\bibliographystyle{aasjournal}

\end{document}